\documentclass[%
reprint,
preprintnumbers,
nofootinbib,
 amsmath,amssymb,
 aps,
prd,
raggedbottom
]{revtex4-1}

\usepackage[]{putexModified}
\usepackage[]{hyperref}
\usepackage[percent]{overpic}

\makeatletter
\DeclareRobustCommand*{\bfseries}{%
  \not@math@alphabet\bfseries\mathbf
  \fontseries\bfdefault\selectfont
  \boldmath
}
\makeatother

\input{glyphtounicode}
\newcommand{\Qq}{\mathbb{Q}_{p^n}}
\newcommand{\Rn}{\mathbb{R}^n}

\begin{document}

\preprint{PUPT-2522}

\title{Geodesic bulk diagrams on the Bruhat--Tits tree}

\author{Steven S. Gubser}%
  \email{ssgubser@princeton.edu}
\author{Sarthak Parikh}%
 \email{sparikh@princeton.edu}
\affiliation{%
  Joseph Henry Laboratories, Princeton University, Princeton, NJ 08544, USA
}%
 
\date{\today}

\begin{abstract}
Geodesic bulk diagrams were recently shown to be the geometric objects which compute global conformal blocks. We show that this duality continues to hold in $p$-adic AdS/CFT, where the bulk is replaced by the Bruhat--Tits tree, an infinite regular graph with no cycles, and the boundary is described by $p$-adic numbers, rather than reals. We apply the duality to evaluate the four-point function of scalar operators of generic dimensions using tree-level bulk diagrams.  Relative to standard results from the literature, we find intriguing similarities as well as significant simplifications. Notably, all derivatives disappear in the conformal block decomposition of the four-point function. On the other hand, numerical coefficients in the four-point function as well as the structure constants take surprisingly universal forms, applicable to both the reals and the $p$-adics when expressed in terms of local zeta functions. Finally, we present a minimal bulk action with nearest neighbor interactions on the Bruhat--Tits tree, which reproduces the two-, three-, and four-point functions of a free boundary theory.
\end{abstract}

\maketitle


\section{Introduction}

Operator product expansions (OPEs) in CFTs defined on $p$-adic spaces were first studied in~\cite{Melzer1989} in the context of two-dimensional CFTs. The $p$-adic OPE is much simpler than its Archimedean\footnote{We will use the term Archimedean to refer to any construction based entirely on real numbers rather than $p$-adic numbers.  So for example, Archimedean field theories are the ones usually studied in textbooks, based on fields defined on $\mathbb{R}^n$ or $\mathbb{R}^{n-1,1}$.  To be precise, the Archimedean property of real numbers is that if $0<|a|<|b|$, then there is some integer $n$ such that $|na|>|b|$.  The failure of the $p$-adic numbers to share this property is generally recognized to be their most fundamental difference from the reals.} counterpart since it does not admit derivative expansions in any obvious way, due to the lack of a notion of local derivatives of functions whose domain is the $p$-adics and whose range is the reals. Consequently, no descendants appear in the $p$-adic OPE; in fact in such a CFT a local stress-tensor is absent. Nevertheless, $p$-adic field theories share several common features with usual field theories, including renormalization group flows and the existence of a Wilson-Fisher fixed point~\cite{Lerner:1989ty,Lerner:1994th,Missarov:2006in,Missarov:2006iu,Gubser:2017vgc}. At the fixed point, anomalous dimensions admit universal expressions independent of the choice of space, whether $p$-adic or real~\cite{Gubser:2017vgc}. 

On the holographic side, the dual geometry permits a natural description in terms of discrete graphs, the simplest of which is the Bruhat--Tits tree --- an infinite regular graph with no cycles and a coordination number $p^n+1$, where $p^n$ is a power of a prime --- that serves as the $p$-adic analog of Euclidean AdS space~\cite{Gubser:2016guj, Heydeman:2016ldy,Gubser:2016htz}.\footnote{See \cite{Heydeman:2016ldy,Bhattacharyya:2017aly} for discussions on connections between tensor networks and the Bruhat--Tits tree.} 
The boundary of the Bruhat--Tits tree is described in terms of the extended $p$-adic numbers $\Qq$ (more precisely, the projective line $\mathbb{P}^1(\Qq)$), which may be viewed as an $n$-dimensional vector space over the unextended $p$-adic numbers $\mathbb{Q}_p$ in much the same way as $\mathbb{C}$ can be viewed as a two-dimensional vector space over $\mathbb{R}$.\footnote{For a quick review of $p$-adic numbers and their extensions, see for example~\cite{Gubser:2016guj}.}
In order to make contact with the standard results in $\Rn$,\footnote{We will sometimes refer to the usual AdS/CFT correspondence over the reals as the Archimedean case or simply as $\Rn$, and we will refer to the non-Archimedean case over the $p$-adics simply as $\Qq$.} we restrict ourselves to a bulk scalar field on a fixed AdS background, since in this case the dual CFT in $\Rn$ lacks a stress tensor as well.\footnote{In~\cite{Gubser:2016htz}, in the context of $p$-adic AdS/CFT, a ``stress-tensor'' like operator dual to ``graviton fluctuations'' was considered. The bulk description consisted of a scalar field defined on the vertices of the infinite Bruhat--Tits tree, with the ``graviton'' described as edge-length fluctuations. The operator dual to edge-length fluctuations shared some properties with the stress-tensor, such as the correct scaling dimension, but seemed to lack others, such as a notion of spin. In this paper, we avoid the open question of the properties of a stress tensor in $p$-adic CFTs by restricting to a fixed bulk without dynamical gravity. For unrelated reasons, CFTs in $\Rn$ without a stress tensor were also recently considered in \cite{Paulos:2016fap,Aharony:2016dwx,Behan:2017emf}.}
 In the $p$-adics, the theory of a bulk scalar on a fixed AdS background is described by a scalar degree of freedom defined on the vertices of the Bruhat--Tits tree. 
Such theories have been previously studied  in the case of a massless scalar~\cite{Zabrodin:1988ep}, massive scalar~\cite{Gubser:2016guj,Heydeman:2016ldy}, as well as an interacting scalar with cubic and quartic interaction vertices~\cite{Gubser:2016guj}.

In \cite{Gubser:2016guj}, a discrete bulk Euclidean action on the Bruhat--Tits tree $T_{p^n}$  was considered,
\eqn{Ssingle}{
S[\phi] &=  \sum_{\langle ab \rangle} {1 \over 2} (\phi_a - \phi_b)^2 \cr &+ \sum_{a \in T_{p^n}} \!\left( {1\over 2} m_{\Delta}^2 \phi_a^2 + {g_3 \over 3!} \phi_a^3 + {g_4 \over 4!} \phi_a^4 \right),
}
which describes a single scalar field $\phi$ living on the vertices. The first sum in \eno{Ssingle} runs over all edges, and the second sum runs over all the vertices of $T_{p^n}$. Here $g_3$ and $g_4$ are coupling constants, and the mass of the scalar is related to the scaling dimension of the dual operator, $\Delta$ via~\cite{Gubser:2016guj}
\eqnQq{mDeltaQq}{
m_{\Delta}^2 \equiv m_{\Qq, \Delta}^2 = {-1 \over \zeta(-\Delta) \zeta(\Delta-n)}\,.
}
The somewhat non-standard notation used in \eno{mDeltaQq} and the rest of the paper is explained at the end of this section. It is worth noting here that the zeta function that appears in \eno{mDeltaQq} is {\it not} the Riemann zeta function; instead, we reserve the symbol $\zeta$ to stand for the so-called  local zeta functions, defined in \eno{zetaQq}-\eno{zetaRn}.

The action in \eno{Ssingle} was used to compute the bulk-to-bulk and the bulk-to-boundary propagators~\cite{Gubser:2016guj,Heydeman:2016ldy}, which in turn yielded the holographic two- and three-point functions, as well as the connected four-point function at tree-level for four identical scalar operators~\cite{Gubser:2016guj}. 
The four-point function, which was restricted to a contact interaction arising from a $\phi^4$ bulk vertex, was found to have a particularly simple form compared to its expression in the usual Archimedean AdS/CFT over $\Rn$.
Notwithstanding, surprising similarities  were discovered between the $p$-adic and Archimedean correlators~\cite{Gubser:2016guj}, with the holographically obtained structure constants admitting universal expressions independent of the field under consideration ($p$-adics or the reals),  hinting at some version of an adelic principle at play~\cite{Freund:1987ck,Brekke:1988yb,Melzer1989}.
The first goal of this paper is to present the full $p$-adic four-point function of scalar composite operators  of arbitrary dimensions, including this time the exchange interactions arising from cubic scalar couplings in the bulk.  Compared to the situation in Archimedean AdS/CFT, the $p$-adic contact and exchange diagrams in the direct and crossed channels have much simpler expressions, and we are able to obtain explicit closed-form expressions for them. These remarkable simplifications arise due to the fact that only the simplest double-trace operators appear in the conformal block decomposition of the diagrams --- the ones with {\it no} derivatives;  additionally no descendants contribute either.  Despite the simplifications, one continues to find striking similarities between the results in $\Rn$ and $\Qq$, and we emphasize these in this paper. 

To evaluate the exchange interactions in various channels, it is useful to employ the geodesic bulk diagram techniques introduced in \cite{Hijano:2015zsa}, often termed geodesic Witten diagrams since they are a simplification of the diagrammatic techniques introduced in \cite{Witten:1998qj}. Roughly speaking, geodesic bulk diagrams are bulk exchange diagrams with the bulk points of integration restricted to the geodesics joining the boundary points, rather than the whole of AdS space. The geodesic bulk diagram so constructed turns out to be directly related to the conformal blocks in the boundary CFT. It is natural to expect analogous relations to hold in $p$-adic AdS/CFT as well, especially because geodesics feature prominently on the Bruhat--Tits tree --- in fact, all paths on the tree are geodesics provided backtracking is disallowed. Indeed, we will show that the relation between geodesic bulk diagrams and conformal blocks carries over to the $p$-adics, and in fact turns out to be both similar to as well as simpler than its Archimedean counterpart. In the process, we obtain the decomposition of geodesic bulk diagrams in both direct as well as crossed channels. 

The final goal of this paper is to construct a bulk action which reproduces the two-, three- and four-point functions of a free-field theory. We do that by introducing a  nearest neighbor interaction in the bulk, evaluating its contributions to the four-point correlator, and tuning the cubic and quartic bulk couplings to arrive at the free-field connected four-point function.

A cautionary word on the non-standard notation used in this paper is in order.  Equations applicable only in $\Rn$ or $\Qq$ will be marked to indicate so. For instance, the mass - scaling dimension relation between mass of a scalar field, and the scaling dimension of the dual operator in $\Qq$ is given by \eno{mDeltaQq}, while in $\Rn$ it is the well known relation
 \eqnRn{mDeltaRn}{
 m_{\Delta}^2 \equiv m_{\mathbb{R}^n, \Delta}^2 = \Delta (\Delta - n)\,.
 }
Throughout the paper, we will be expressing results in terms of the local zeta functions, defined to be
  \eqnQq{zetaQq}{
 \zeta(s) \equiv \zeta_{\mathbb{Q}_{p}}(s) = {1 \over 1-p^{-s}}
 }
and
 \eqnRn{zetaRn}{
 \zeta(s) \equiv \zeta_{\mathbb{R}}(s) = \pi^{-s/2}\Gamma_{\rm Euler}(s/2)\,,
 }
in $\Qq$ and $\Rn$, respectively.\footnote{As in \cite{Gubser:2017vgc}, we avoid defining local zeta functions for $\mathbb{R}^n$ or $\mathbb{Q}_{p^n}$ and prefer to use only the local zeta functions defined in \eno{zetaQq} and \eno{zetaRn}.} The reason for defining the local zeta functions is that holographic correlators in $\Rn$ and $\Qq$ have more or less a universal form, when expressed in terms of these local zeta functions.\footnote{From a purely CFT perspective, it was shown in~\cite{Gubser:2017vgc} (in the context of the $p$-adic $O(N)$ model), that correlators and anomalous dimensions of various operators admit expressions expressible entirely in terms of suitably defined local gamma and beta functions, which are ultimately constructed out of the local zeta functions defined in \eno{zetaQq}-\eno{zetaRn}. It is satisfying to see the same functions appear both from purely bulk calculations as well as purely boundary considerations, especially since in the context of $p$-adic AdS/CFT the bulk/boundary correspondence relates two seemingly disparate constructs: a bulk described by a discrete tree and a boundary described by the continuum of $p$-adic numbers. We will not have occasion to use the local gamma and beta functions (though \eno{betaDef} is related); however, to prevent confusion we will always use Euler's gamma and beta functions as $\Gamma_{\rm Euler}(z)$ and ${\rm B}_{\rm Euler}(z,w)$.}

Equations which hold both in $\mathbb{R}^n$ and $\mathbb{Q}_{p^n}$ will be left unmarked. For instance in $\Rn$ and $\Qq$, it will be useful to define\footnote{In $\mathbb{R}^n$, $\beta^{(\Delta_1,\Delta_2)}$ reduces to the usual Euler beta function, $\beta^{(\Delta_1,\Delta_2)}_{\mathbb{R}^n} = {\rm B}_{\rm Euler}(\Delta_1/2,\Delta_2/2)$.}
 \eqn{betaDef}{
 \beta^{(\Delta_1, \Delta_2)} \equiv {\zeta(\Delta_1)\zeta(\Delta_2) \over \zeta(\Delta_1 + \Delta_2)}\,.
 }

The organisation of the rest of the paper is as follows. In section \ref{GBD} we establish the relation between $p$-adic conformal blocks and geodesic bulk diagrams on the Bruhat--Tits tree, and in section \ref{FOURPT} we use results from section \ref{GBD} as well as some bulk propagator identities to evaluate the full four-point function of scalar operators (including contact interactions as well as exchange diagrams in all channels). In section \ref{FREECFT} we describe a minimal bulk construction which yields the two-, three-, and four-point functions of a free theory on the boundary. We end with a summary and discussion of some open questions in section \ref{DISCUSSION}. In the appendices we list some additional propagator identities on the Bruhat--Tits tree, illustrate crossing symmetry in the $p$-adics, and explore the connection between nearest neighbor interactions and derivative couplings.

\section{Conformal blocks and geodesic bulk diagrams}
\label{GBD}
\subsection{The $p$-adic OPE and the three-point function}
\label{PADICOPE}
 The OPE between two (scalar) operators in a CFT takes the general form
 \eqn{OPE}{
\mathcal{O}_1(x_1) \mathcal{O}_2(x_2) = \sum_{r} \widetilde{C}_{12r} |x_{12}|^{-\Delta_1-\Delta_2+\Delta_r} \mathcal{O}_r(x_2)\,,
 }
where the sum is over all operators in the CFT. The operator ${\cal O}_r$ has scaling dimension $\Delta_r$, and $\widetilde C_{ijk}$s are the OPE coefficients. In a $p$-adic CFT, descendants do not appear in the OPE, and the index $r$ runs only over the `primaries'~\cite{Melzer1989}. A scalar `primary' operator $\mathcal{O}$ with scaling dimension $\Delta$ transforms under $p$-adic conformal transformations,
 \eqnQq{LFT}{
z \rightarrow z^\prime &= {az + b \over cz + d}\,,\;\;\; \begin{pmatrix}
a & b \cr c & d
\end{pmatrix} \in {\rm PGL}(2,\Qq)
 }
as
 \eqnQq{PrimaryPadic}{
 \mathcal{O}^\prime(z^\prime) &=   \left| {ad -bc \over (cz+d)^2}\right|^{-\Delta} \mathcal{O}(z)\,,
 }
 where $|\cdot|$ represents the $p$-adic norm.
 This serves as the defining property of scalar `primary' operators in $p$-adic CFTs~\cite{Melzer1989}. 
 Moreover, we postulate orthonormality
 \eqn{TwoPtIJ}{
\langle \mathcal{O}_i(x_1) \mathcal{O}_j(x_2) \rangle = {\delta_{ij} \over |x_{12}|^{2\Delta_j} }\,.
}

First consider a theory of bulk scalars of generic masses with cubic couplings of the form $\phi_i \phi_j \phi_k$ in a fixed AdS background.
The $p$-adic OPE of two non-degenerate operators takes the form in \eno{OPE} where the sum runs over all operators, including multi-trace. 
Inserting a single-trace operator $\mathcal{O}_3$ at $x_3$ of dimension $\Delta_3 \neq \Delta_1 + \Delta_2$ such that $|x_{12}| < |x_{13}|, |x_{23}|$, the three-point function of three single-trace operators following from the OPE is
\eqnQq{ThreePtOPE}{
& \langle \mathcal{O}_1(x_1) \mathcal{O}_2(x_2) \mathcal{O}_3(x_3)\rangle \cr
 &= \sum_{r} \widetilde{C}_{12r} |x_{12}|^{-\Delta_1-\Delta_2+\Delta_r} \langle \mathcal{O}_r(x_2) \mathcal{O}_3(x_3) \rangle \cr
 &=   \widetilde{C}_{123}  |x_{12}|^{-\Delta_1-\Delta_2+\Delta_3} |x_{23}|^{-2\Delta_3}\,,
}
where we used the orthonormality property \eno{TwoPtIJ} and assumed that the cubic coupling $\phi_1 \phi_2 \phi_3$ is present.  Since $p$-adic conformal invariance fixes the form of the three-point function up to an overall constant $\widetilde{f}_{ijk}$~\cite{Melzer1989}, we may set \eno{ThreePtOPE} to
\eqn{ThreePtConfInv}{
   {\widetilde{f}_{123} \over |x_{12}|^{\Delta_1+\Delta_2-\Delta_3} |x_{23}|^{\Delta_2+\Delta_3-\Delta_1} |x_{13}|^{\Delta_3+\Delta_1-\Delta_2}}\,.
}
Ultrametricity of the $p$-adic norm implies $|x_{13}| = |x_{23}|$.\footnote{\label{Ultrametricity}The proof proceeds as follows. Rewrite $|x_{13}| = |x_{12} + x_{23}|$. By assumption, $|x_{12}| < |x_{23}|$. Then the desired relation follows directly from the general property of the $p$-adic norm that $|x+y| = |x|$ if $|x|>|y|$.} This immediately yields at leading order (i.e.~tree-level),
\eqnQq{Cfijk}{
\widetilde{C}_{123} = \widetilde{f}_{123} \,.
}

The OPE coefficient can be determined by working out the three-point function holographically. For operators with generic dimensions $\Delta_1, \Delta_2, \Delta_3$, which are bulk duals to $\phi_1, \phi_2, \phi_3$ appearing in a cubic vertex,\footnote{In the special case of non-generic scaling dimensions with $\Delta_i + \Delta_j  - \Delta_k = 0$, anomalous dimensions become important at tree-level. We will not address this case here.} the standard prescription in $\Rn$ to compute the tree-level contribution to the three-point function is to evaluate the integral\footnote{The normalization differs slightly from the one used in \cite{Gubser:2016guj}, due to the different choice of normalization for the two-point function. (See also footnote \ref{fn:translate}.)}
 \eqnRn{ThreePtRn}{
 \langle \mathcal{O}_1(\vec{x}_1)& \mathcal{O}_2(\vec{x}_2) \mathcal{O}_3(\vec{x}_3) \rangle \cr =\; &  \mathcal{N}_3 \int {d^{n+1} y \over y_0^{n+1}} \left(\prod_i^3 \hat{K}_{\Delta_i}(y_0, \vec{y}-\vec{x}_i)\!\right)\!,
}
where $\hat{K}_\Delta$ is the unnormalized bulk-to-boundary propagator
 \eqnRn{KhatRn}{
 \hat{K}_\Delta(y_0,\vec{y}-\vec{x}) = {y_0^\Delta \over (y_0^2 + (\vec{y}-\vec{x})^2)^{\Delta}}\,,
}
and
 \eqn{NKDef}{
 \mathcal{N}_3 \equiv -g_3 \left(\prod_i^3 \sqrt{\widetilde{c}_{\Delta_i}}\right) \,,
 }
with
 \eqnRn{ctDeltaRn}{
 \widetilde{c}_{\Delta} \equiv {c_{\Delta} / (2\Delta-n)}\,,
 }
and
  \eqn{cDelta}{
 c_{\Delta} \equiv {\zeta(2\Delta) / \zeta(2\Delta-n)}\,.
 }

In $\Qq$, integration over the bulk point $y$ gets replaced by a sum over all vertices $a$ of the Bruhat--Tits tree~\cite{Gubser:2016guj}, giving
 \eqnQq{ThreePtQq}{
 \langle \mathcal{O}_1(x_1)& \mathcal{O}_2(x_2) \mathcal{O}_3(x_3) \rangle \cr &=\; \mathcal{N}_3 \sum_{a \in T_{p^n}} \left(\prod_i^3 \hat{K}_{\Delta_i}(a,x_i)\right),
}
where the unnormalized bulk-to-boundary propagator $\hat{K}_\Delta$ is given by
 \eqnQq{KhatQq}{
 \hat{K}_\Delta(a,x) = \hat{K}_\Delta(y_0,y-x) = {|y_0|^\Delta \over |y_0, y-x|_s^{2\Delta}},
 }
 $\mathcal{N}_3$ is as in \eno{NKDef}, $\widetilde{c}_\Delta$ is defined to be
 \eqnQq{ctDeltaQq}{
 \widetilde{c}_\Delta \equiv {c_{\Delta} \over p^{\Delta}/\zeta(2\Delta-n)}\,,
 }
and $c_\Delta$ is given by \eno{cDelta}. 
In the first equality in \eno{KhatQq}, the bulk vertex $a$ is re-expressed in terms of the boundary coordinate $y \in \Qq$ and the bulk depth coordinate $y_0 \in p^\mathbb{Z}$ which together specify $a$~\cite{Gubser:2016guj}. The notation $|z,w|_s$ stands for supremum norm, $|z,w|_s \equiv \sup \{ |z|, |w| \}$.\footnote{For brevity, from now on we will suppress the vector symbol on boundary coordinates in $\Rn$, and analogous to the $p$-adics, refer to bulk coordinates with lower-case Latin alphabets such as $a,b,c$, so that for example, $a = (y_0, \vec{y})$. Boundary coordinates will usually be denoted using letters from the other end of the alphabet, for example $x, y, z$.}

 In $\Rn$, evaluating the three-point function given in \eno{ThreePtRn}, leads to \eno{ThreePtConfInv} with~\cite{Freedman:1998tz}
 \eqnRn{StrConstRn}{
\widetilde{f}_{123}   &= {\mathcal{N}_3\over 2} f_{123}\,,
}
where
 \eqn{fijk}{
 f_{ijk} &\equiv \zeta(\Delta_i+\Delta_j+\Delta_k - n) \cr 
 &\times { \zeta(\Delta_i+\Delta_{jk}) \zeta(\Delta_j + \Delta_{ki}) \zeta(\Delta_k + \Delta_{ij})  \over \zeta(2\Delta_i)\zeta(2\Delta_j)\zeta(2\Delta_k)}\,.
 }
Here $\Delta_{ij} = \Delta_i - \Delta_j$ and $\zeta(s)$ is defined in \eno{zetaRn}. By comparison in $p$-adic AdS/CFT, to evaluate \eno{ThreePtQq} a simple generalization of the computation of the holographic three-point amplitude presented in~\cite{Gubser:2016guj} leads to \eno{ThreePtConfInv} with
 \eqnQq{StrConstQq}{
\widetilde{f}_{123} = \mathcal{N}_3\: f_{123}\,,
}
where $f_{123}$ is again given by \eno{fijk}, and $\zeta(s)$ is defined in \eno{zetaQq}.

While the equalities in \eno{StrConstRn} and \eno{StrConstQq} are sufficient to show the striking similarity between the form of the structure constants in $\Rn$ and $\Qq$, an alternate form for the structure constants will be more instructive when we later compare the scalar four-point functions in $\Rn$ and $\Qq$. For $n=2$ in $\Rn$,
 \eqnRn{StrConstRnAlt}{
\widetilde{f}_{123}  &\stackrel{(n=2)}{=}  {-\mathcal{N}_3 \over \widetilde{c}_{\Delta_1+\Delta_2}} \beta^{(\Delta_3 + \Delta_{12},\Delta_3 - \Delta_{12})} \cr
 & \quad \times \sum_{M=0}^\infty {a_{M}^{(\Delta_1,\Delta_2)} \over m_{\Delta_3}^2 - m_{\Delta_1+\Delta_2 + 2M}^2}\,,
}
where $\widetilde{c}_\Delta$ is defined in \eno{ctDeltaRn}, $\beta^{(s,t)}$ is defined in \eno{betaDef}, the mass squared in $\mathbb{R}^n$ is given by \eno{mDeltaRn}, and
 \eqnRn{aDefRn}{
a_M^{(\Delta_1, \Delta_2)} &\equiv {1 \over \beta^{(2\Delta_1 + 2M, 2\Delta_2 + 2M)}} {(-1)^M \over M!}\cr 
&\times {(\Delta_1)_M (\Delta_2)_M \over \left(\Delta_1 + \Delta_2 + M - n/2 \right)_M}\,,
}
where $(\Delta)_M \equiv \Gamma_{\rm Euler}(\Delta+M) / \Gamma_{\rm Euler}(\Delta)$ is the Pochhammer symbol. On the other hand, in $\Qq$ for general $n$,
 \eqnQq{StrConstQqAlt}{
\widetilde{f}_{123} &= {-\mathcal{N}_3  \over \widetilde{c}_{\Delta_1+\Delta_2}}  \beta^{(\Delta_3+\Delta_{12},\Delta_3-\Delta_{12})} \cr 
&\times {a^{(\Delta_1,\Delta_2)} \over m_{\Delta_3}^2 - m_{\Delta_1+\Delta_2}^2} \,.
}
Here $\widetilde{c}_\Delta$ is defined in \eno{ctDeltaQq} and $\beta^{(s,t)}$ is defined in \eno{betaDef}. The $p$-adic mass squared is given by \eno{mDeltaQq}, and
 \eqnQq{aDefQq}{
 a^{(\Delta_1,\Delta_2)} \equiv {1 \over \beta^{(2\Delta_1, 2\Delta_2)}}\,.
 }
 Although the equality in \eno{StrConstRnAlt} holds only for $n=2$, the comparison between \eno{StrConstRnAlt} and \eno{StrConstQqAlt} proves useful in section \ref{FOURPT}, when we compare the four-point functions in $\Rn$ and $\Qq$ for any $n$.
 The absence of an infinite sum in \eno{StrConstQqAlt} along the lines of \eno{StrConstRnAlt} turns out to be directly related to the absence of derivatives in the OPE in $p$-adic CFTs. We will return to this point in section \ref{FOURPT}. It is worth making the trivial observation that at $M=0$, $a_M^{(\Delta_1, \Delta_2)}$ in \eno{aDefRn} reduces to the simple form
 \eqnRn{aMZero}{
  a^{(\Delta_1, \Delta_2)}_{0} = {1 \over \beta^{(2\Delta_1, 2\Delta_2 )}},
  }
which is to be compared with the definition \eno{aDefQq} in $\mathbb{Q}_{p^n}$.
 
\subsection{Four-point contact diagram}
 \begin{figure}[t]
  \centerline{\includegraphics[width=2.3in]{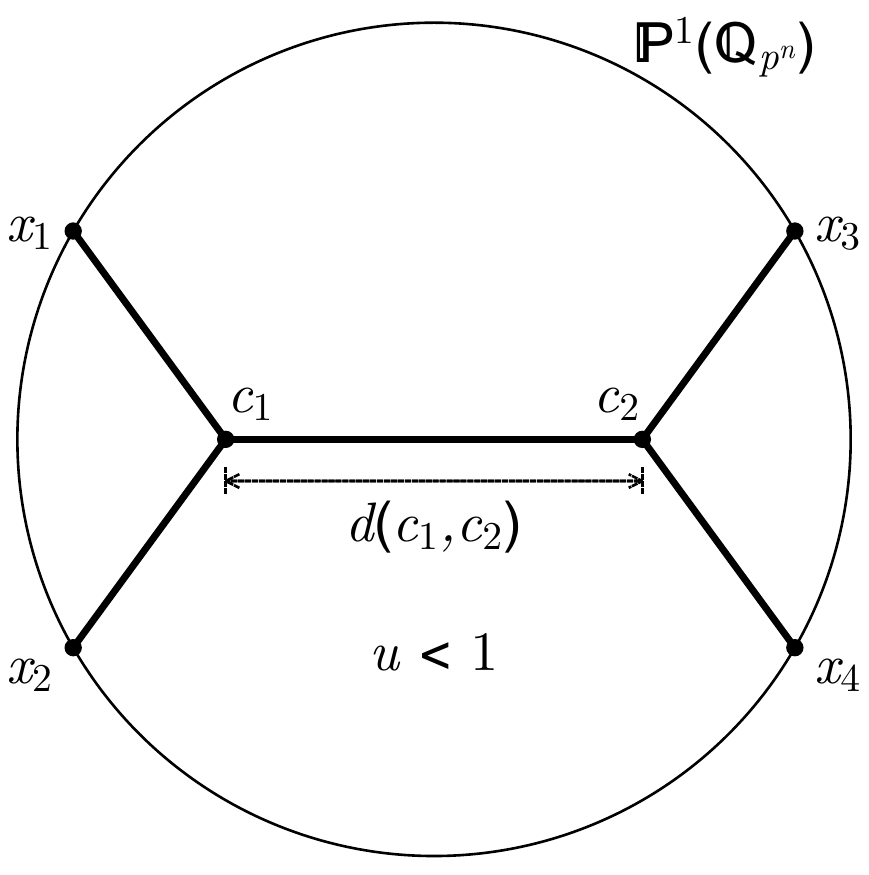}}
  \caption{Boundary points $x_i$ in the s-channel configuration. Solid lines are geodesics on the Bruhat--Tits tree, tracing the path joining the four points together. Bulk points $c_1, c_2$ are uniquely fixed once $x_i$ are specified. The conformal cross-ratio in \eno{uvDef} is given by $u = p^{-d(c_1,c_2)}$ where $d(c_1,c_2)$ is the graph distance between points $c_1$ and $c_2$. In the $s$-channel configuration, $u<1$.}\label{figSchannel}
 \end{figure}
 
 For four boundary points $x_1, x_2, x_3$ and $x_4$, the conformal cross-ratios $u$ and $v$ are defined to be
 \eqnQq{uvDef}{
 u \equiv \left| {x_{12}x_{34} \over x_{13} x_{24}}\right| \qquad v \equiv \left| {x_{14}x_{23} \over x_{13} x_{24}}\right|\,.
 }
In this paper, the  points of insertion of external scalar operators will always be in an `$s$-channel' configuration on the boundary (see figure \ref{figSchannel}). The defining property of `$s$-channel' is that the cross-ratio $u<1$. A striking consequence of ultrametricity is that, $u<1 \Rightarrow v = 1$. To prove this, observe that 
 \eqnQq{vOneProof}{
  {x_{14}x_{23} \over x_{13} x_{24}} = 1- {x_{12}x_{34} \over x_{13} x_{24}}\,.
 }
 The claim then follows directly from an application of the ultrametric property of $p$-adic norms described in footnote~\ref{Ultrametricity}.
 
 We define the $p$-adic four-point contact amplitude to be the sum
\eqnQq{Dfn}{
 \mathcal{D}(x_i) \equiv {1 \over W_0(x_i)} \sum_{a \in T_{p^n}} \left(\prod_{i}^4 \hat{K}_{\Delta_i} (a,x_i)\right)\,,
 }
where
 \eqnQq{WDeltaDef}{
 W_\Delta(x_i) &\equiv \hat{K}_{\Delta_1}(c_1, x_1)  \hat{K}_{\Delta_2}(c_1, x_2) 
  \hat{G}_\Delta(c_1, c_2) \cr
  & \times \hat{K}_{\Delta_3}(c_2, x_3)  \hat{K}_{\Delta_4}(c_2, x_4) \,,
  }
  and $\hat{G}_\Delta$ is the unnormalized bulk-to-bulk propagator, given by~\cite{Gubser:2016guj}
 \eqnQq{GhatQq}{
 \hat{G}_\Delta(c_1,c_2) = p^{-\Delta d(c_1,c_2)} = u^\Delta\,,
 }
  where $c_1$ and $c_2$ are the unique points of intersection of the geodesics joining together the boundary points $x_i$ (see figure \ref{figSchannel}), and $u$ is the cross-ratio defined in \eno{uvDef}. 
The product in \eno{WDeltaDef} evaluates to (see \cite{Gubser:2016guj} for similar computations)
 \eqnQq{WDeltaExpressQq}{
 W_\Delta(x_i) &=  u^\Delta\: W_0(x_i),
 }
 where
  \eqnQq{WZeroExpressQq}{
 W_0 &= {1 \over |x_{12} x_{34}|^{\sigma/2}} \prod_{1 \leq i < j \leq 4} {1 \over |x_{ij}|^{\Delta_i+\Delta_j - \sigma/2}},
 }
 with $\sigma \equiv \sum_{i=1}^4 \Delta_i$. It is clear that $W_0(x_i)$ carries the trivial coordinate dependence of the four-point function. An alternate representation for $W_0$ is 
\eqnQq{WZeroExpressAlt}{
W_0 &= \left|x_{24} \over x_{14} \right|^{\Delta_{12}} \left|x_{14} \over x_{13} \right|^{\Delta_{34}}  { v^{(\Delta_{12}-\Delta_{34})/2} \over |x_{12}|^{\Delta_1+\Delta_2}  |x_{34}|^{\Delta_3+\Delta_4}},
}
where we can freely set $v=1$ in the `$s$-channel'.

 The sum over Bruhat--Tits tree in \eno{Dfn} was computed in~\cite{Gubser:2016guj} in the special case of identical $\Delta_i$. Generalizing to non-identical $\Delta_i$, we obtain\footnote{\label{fn:convergence}For convergence of the sum in \eno{Dfn}, we require:
$$
\sum_{i=1}^4\Delta_i > n \qquad \quad \Delta_2 + \Delta_3 + \Delta_4 > \Delta_1  \quad \hbox{ and other permutations.}
$$
The computation proceeds straightforwardly using the tree-summation methods described in \cite{Gubser:2016guj}. Later in section \ref{CONTACT}, we will provide an alternate derivation of \eno{DfnExpress}.
}
\eqnQq{DfnExpress}{
 \mathcal{D}(x_i) = u^{\Delta_A} f_{34A}  +  u^{\Delta_B} f_{12B}\,,
 }
where
\eqn{DeltaAB}{
\Delta_A = \Delta_1 + \Delta_2 \qquad \Delta_B = \Delta_3 + \Delta_4\,,
}
and the $f_{ijk}$s are given by \eno{fijk}. 
We note that $\mathcal{D}(x_i)$ depends on the coordinates only through the cross-ratio $u$, and from here on we will simply write it as $\mathcal{D}(u)$.
The tree-level four-point function is thus given by
\eqnQq{FourPtContact}{
\langle \mathcal{O}_1(x_1) \mathcal{O}_2(x_2) \mathcal{O}_3(x_3) \mathcal{O}_4(x_4) \rangle &= \mathcal{N}_4 W_0(x_i) \mathcal{D}(u),
}
assuming no bulk cubic couplings are present, with $\mathcal{N}_4$ given by \eno{NKDef}. In $\Rn$, conformal invariance constrains the four-point function of scalar operators to be of the form
 \eqnRn{FpfConfInv}{
 \langle \mathcal{O}_1(x_1) \mathcal{O}_2(x_2) \mathcal{O}_3(x_3) \mathcal{O}_4(x_4) \rangle = W_0(x_i) g(u,v),
}
where $W_0(x_i)$ is given by
 \eqnRn{WZeroExpressRn}{
 W_0 &\equiv \left( {|x_{24}| \over |x_{14}|} \right)^{\Delta_{12}} \left( {|x_{14}| \over |x_{13}|} \right)^{\Delta_{34}} \cr
 & \quad \times {1\over |x_{12}|^{\Delta_1+\Delta_2}  |x_{34}|^{\Delta_3+\Delta_4}},
 }
  and $g(u,v)$ is an arbitrary function of cross-ratios $u$ and $v$, defined to be
 \eqnRn{uvDefRn}{
  u \equiv  {|x_{12}||x_{34}| \over| x_{13}| |x_{24}|} \qquad v \equiv  {|x_{14}||x_{23}| \over |x_{13}| |x_{24}|}\,,
 }
 where $|\cdot|$ are $L^2$-norms in $\Rn$.
  As noted above \eno{vOneProof}, in the `$s$-channel' in $\Qq$ one of the cross-ratios is trivial.  So we see that up to an overall normalization factor, $\mathcal{D}(u)$ is the $p$-adic analog of $g(u,v)$.  From here on we will not concern ourselves with the four-point function $\langle \mathcal{O}\ldots \mathcal{O}\rangle$, but study directly the amplitude $\mathcal{D}(u)$, which is stripped off of the trivial kinematic factors and contains only the dynamical information of the theory.

\subsection{$p$-adic conformal blocks}
\label{CONFWAVE}

 In analogy with the decomposition of $g(u,v)$ into conformal blocks in $\Rn$, the amplitude $\mathcal{D}(u)$ may also be decomposed into (scalar) conformal blocks $\mathcal{G}_{\Delta}(u)$,
\eqnQq{DfnExpand}{
\mathcal{D}(u) = \sum_{r} C_{12r} C_{34r}\: \mathcal{G}_{\Delta_r}(u)\,.
}
Comparing with \eno{DfnExpress}, we see the $p$-adic conformal blocks are simply given by
\eqnQq{ConformalBlock}{
\mathcal{G}_{\Delta}(u) = u^{\Delta}\,.
}
We can now also identify the kinematic factor $W_\Delta(x_i)$ in \eno{WDeltaExpressQq} with the scalar conformal partial wave, since
 \eqnQq{WDeltaAgain}{
 W_\Delta(x_i) = W_0(x_i) \mathcal{G}_\Delta(u)\,.
 }
 Incidentally in $\Rn$, the conformal partial wave takes the form
  \eqnRn{WDeltaRn}{
 W_\Delta(x_i) \equiv W_0(x_i) \mathcal{G}_\Delta(u,v)\,,
 }
 where ${\cal G}(u,v)$ is the scalar conformal block.\footnote{The contrast  between the $p$-adic conformal blocks \eno{ConformalBlock} and  scalar conformal blocks in $\Rn$ is striking. In $\Rn$ the conformal block admits a double power series expansion in $u^2$ and $(1-v^2)$,
\eqnRn{ConformalBlocksRn}{
\mathcal{G}_{\Delta}(u,v) = u^{\Delta}  \sum_{m,n=0}^\infty a_{mn} u^{2m} (1-v^2)^n\,,
}
for some (known) coefficients $a_{mn}$~\cite{Dolan:2000ut}.}

To arrive at \eno{DfnExpand} starting from the OPE, consider a bulk theory of four scalar fields $\phi_i$ with a quartic interaction of the form $\phi_1 \phi_2 \phi_3 \phi_4$ and no cubic coupling. The OPEs to consider are
\eqn{TwoOPEs}{
\mathcal{O}_1(x_1) \mathcal{O}_2(x_2) &= \sum_{r} \widetilde{C}_{12r} |x_{12}|^{-\Delta_1-\Delta_2+\Delta_r} \mathcal{O}_r(x_2) \cr
\mathcal{O}_3(x_3) \mathcal{O}_4(x_4) &= \sum_{r} \widetilde{C}_{34r} |x_{34}|^{-\Delta_3-\Delta_4+\Delta_r} \mathcal{O}_r(x_4)\,.
}
For the OPEs to make sense, we must have
\eqn{OPEradius}{
|x_{12}|, |x_{34}| < |x_{13}|, |x_{24}|,|x_{14}|, |x_{23}|\,.
}
These requirements are consistent with the $s$-channel configuration shown in figure \ref{figSchannel}. In fact the conditions \eno{OPEradius} are stronger than just requiring $u<1$, and due to ultrametricity, they lead to the following equalities:
\eqnQq{vStronger}{
|x_{13}| = |x_{24}| = |x_{14}| = |x_{23}|\,,
}
which are consistent with but stronger than $v=1$. Then using \eno{TwoPtIJ} we obtain
\eqn{FourPtOPE}{
& \langle \mathcal{O}_1(x_1) \mathcal{O}_2(x_2) \mathcal{O}_3(x_3) \mathcal{O}_4(x_4) \rangle \cr &=  \sum_{r} \widetilde{C}_{12r}  \widetilde{C}_{34r} |x_{12}|^{-\Delta_1-\Delta_2+\Delta_r} |x_{34}|^{-\Delta_3-\Delta_4+\Delta_r} \cr 
& \qquad \quad\times |x_{24}|^{-2\Delta_r}\,.
}
Recalling the properties of the $p$-adic OPE from section \ref{PADICOPE}, and exploiting the fact that there are no bulk cubic couplings but only the quartic coupling $\phi_1\phi_2\phi_3\phi_4$, we conclude that at tree-level, the index $r$ in \eno{TwoOPEs} runs  over only the double-trace operators $\mathcal{O}_B \equiv \mathcal{O}_3 \mathcal{O}_4$ in the first line, and $\mathcal{O}_A \equiv \mathcal{O}_1 \mathcal{O}_2$ in the second line of \eno{TwoOPEs}, with $\Delta_A$ and $\Delta_B$ given by \eno{DeltaAB} to leading order.
 Then using \eno{FourPtContact} and various equalities from \eno{vStronger}, it is easy to show that \eno{FourPtOPE} reproduces \eno{DfnExpress} provided we make the identification
 \eqnQq{Ctildes}{
 \widetilde{C}_{12r}\widetilde{C}_{34r} = \mathcal{N}_4 C_{12r} C_{34r} \qquad r=A,B
 }
with
\eqnQq{OPEcoeffsProd}{
C_{12A} C_{34A} &=  f_{34A} \cr
C_{12B} C_{34B} &=  f_{12B}\,,
}
where $f_{ijk}$s are given in \eno{fijk}.
Crossing symmetry imposes the $p$-adic OPE coefficients $\widetilde{C}_{ijk}$ to satisfy associativity:\cite{Melzer1989}
\eqnQq{OPEAssoc}{
\sum_r \widetilde{C}_{ijr} \widetilde{C}_{k\ell r} = \sum_r \widetilde{C}_{i \ell r} \widetilde{C}_{jkr} = \sum_r \widetilde{C}_{ikr} \widetilde{C}_{j \ell r}\,.
}
The coefficients given in \eno{Ctildes}-\eno{OPEcoeffsProd} satisfy \eno{OPEAssoc} at leading order in the coupling. 
The associativity constraints with $\{i,j,k,\ell\}$ being some permutation of $\{1,2,3,4\}$ leave some freedom to rescale the $C_{ijk}$ while maintaining \eno{OPEcoeffsProd}.\footnote{The aforementioned associativity boils down to verifying the unobvious identity
$$ f_{34A} + f_{12B} = f_{24C} + f_{13D} = f_{23E} + f_{14F}  \,,$$
where the $f_{ijk}$s are given in \eno{fijk}, $\Delta_A, \Delta_B$ are given in \eno{DeltaAB} and 
$\Delta_C = \Delta_1 + \Delta_3, \Delta_D = \Delta_2 + \Delta_4, \Delta_E = \Delta_1 + \Delta_4$ and $\Delta_F = \Delta_2 + \Delta_3$. \label{fn:assoc}
}

\subsection{Geodesic bulk diagrams}
\label{GEODESICDIAG}

\begin{figure}[t]
  \centerline{\includegraphics[width=2.3in]{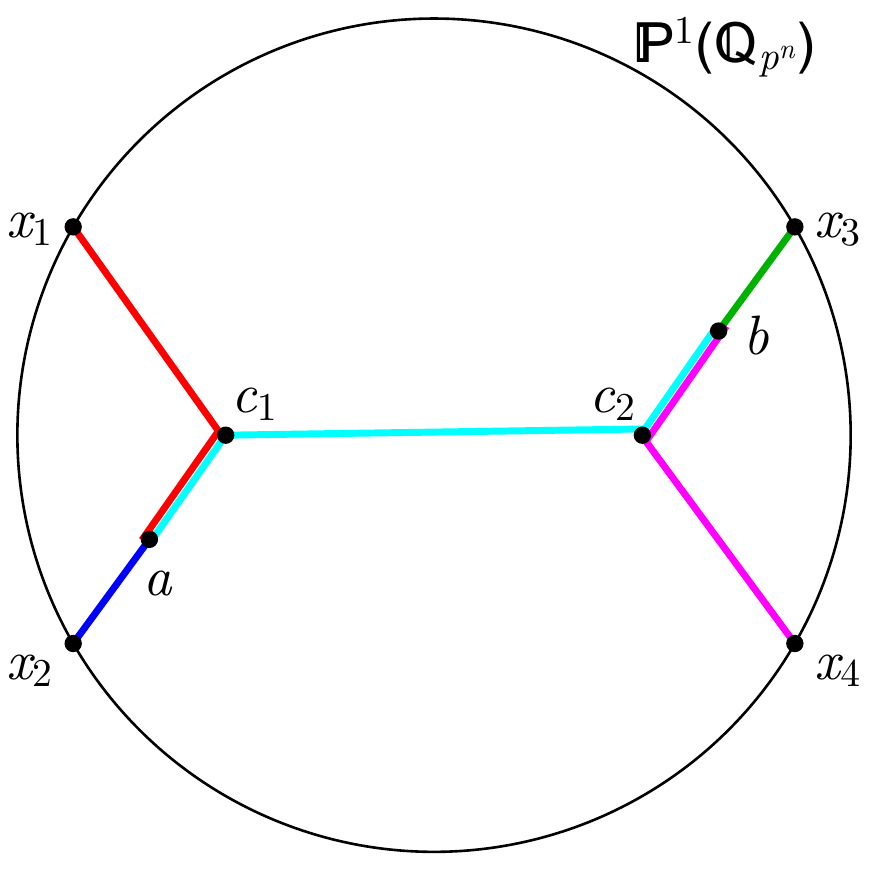}}
  \caption{(Color online.) A geodesic subway diagram. Bulk point $a$ runs along the geodesic joining $x_1$ with $x_2$, and $b$ runs along the geodesic joining $x_3$ with $x_4$. Colors differentiate the individual propagators in \eno{GeoBulkDiag}.}\label{figGeodesic}
 \end{figure}

Following \cite{Hijano:2015zsa}, we define the geodesic bulk diagram $\mathcal{W}_{\Delta}^S$ to be
 \eqnQq{GeoBulkDiag}{
 \mathcal{W}_\Delta^S &\equiv \sum_{a \in \gamma_{12}} \sum_{b \in \gamma_{34}} \!\!\left( \hat{K}_{\Delta_1}(x_1,a)\hat{K}_{\Delta_2}(x_2,a) \right. \cr 
 & \left. \quad \times\: \hat{G}_{\Delta}(a,b) \hat{K}_{\Delta_3}(b,x_3)\hat{K}_{\Delta_4}(b,x_4) \right),
 }
where the bulk point $a$ ($b$) is summed over the unique bulk geodesic $\gamma_{12}$ ($\gamma_{34}$) on the Bruhat--Tits tree joining boundary points $x_1$ and $x_2$ ($x_3$ and $x_4$). The label $S$ is unrelated to the `$s$-channel' configuration of the boundary points $x_i$, but indicates that the bulk points $a$ and $b$ are integrated along $\gamma_{12}$ and $\gamma_{34}$, respectively. Later in section \ref{TUEXCHANGE} we will have occasion to define ${\cal W}_\Delta^T$ and ${\cal W}_{\Delta}^U$. Figure \ref{figGeodesic} shows the {\it subway} diagram, i.e.~a Feynman diagram on the Bruhat--Tits tree, for the geodesic bulk diagram ${\cal W}_\Delta^S$.  Summing over $a$ and $b$ as indicated in \eno{GeoBulkDiag} leads immediately to (see appendix \ref{MOREDETAILS} for details)
\eqnQq{GeodesicConfWaveQq}{
{\mathcal{W}_{\Delta}^S \over W_{\Delta}}= \beta^{(\Delta + \Delta_{12},\Delta - \Delta_{12})}\beta^{(\Delta+\Delta_{34},\Delta-\Delta_{34})}.
}
The result \eno{GeodesicConfWaveQq} is to be compared with results of \cite{Hijano:2015zsa}, where it is shown that that the Archimedean geodesic bulk diagram, ${\cal W}_\Delta^S$ is related to the conformal partial wave $W_\Delta$ via
\eqnRn{GeodesicConfWaveRn}{
{\mathcal{W}_{\Delta}^S \over W_{\Delta}}= {1\over 4}\beta^{(\Delta + \Delta_{12},\Delta - \Delta_{12})}\beta^{(\Delta+\Delta_{34},\Delta-\Delta_{34})}, 
}
and $\beta^{(s,t)}$ is defined in \eno{betaDef}.\footnote{\label{fn:notation}Our notation differs slightly from the one used in \cite{Hijano:2015zsa}. In \cite{Hijano:2015zsa}, 
 $$
 \beta_{\Delta ij} \equiv {1\over 2} \beta^{(\Delta+\Delta_{ij},\Delta-\Delta_{ij})}_{\Rn} = {1\over 2}{\zeta_{\mathbb{R}}(s)\zeta _{\mathbb{R}}(t)\over \zeta_{\mathbb{R}}(s+t)}\,,
 $$
 so that ${\cal W}_\Delta^S = \beta_{\Delta 12} \beta_{\Delta 34} W_\Delta$ (c.f.~(3.3) of \cite{Hijano:2015zsa}).
 The difference in notation is the origin of the explicit factor of $1/4$ in \eno{GeodesicConfWaveRn}.}  Remarkably, comparing \eno{GeodesicConfWaveRn} with \eno{GeodesicConfWaveQq}, we see that the proportionality factors in $\Rn$ and $\Qq$ have an (almost) identical form.

A special case of interest corresponds to setting all external dimensions equal to the dimension of the exchanged scalar. In this case, the tree-sum in \eno{GeoBulkDiag} simplifies to (where now all $\Delta_i = \Delta$)
\eqnQq{GeoBulkDiagDelta}{
\mathcal{W}_{\Delta}^S &= W_0(x_i) \times  \sum_{a \in \gamma_{12}} \sum_{b \in \gamma_{34}} \hat{G}_{\Delta}(a,b),
}
On the other hand, setting all dimensions equal in \eno{GeodesicConfWaveQq} yields
\eqnQq{GeodesicConfWaveDelta}{
{\mathcal{W}_{\Delta}^S \over W_{\Delta}} = \left(\beta^{(\Delta,\Delta)}\right)^2\,.
}
From \eno{GeoBulkDiagDelta}, \eno{GeodesicConfWaveDelta} and \eno{WDeltaAgain}, it follows
\eqnQq{ConfBlockGhatQq}{
\left(\beta^{(\Delta,\Delta)}\right)^2 \mathcal{G}_{\Delta}(u) = \sum_{a \in \gamma_{12}} \sum_{b \in \gamma_{34}} \hat{G}_{\Delta}(a,b)\,.
}
For comparison, the Archimedean analog of \eno{ConfBlockGhatQq} is~\cite{Hijano:2015zsa,Czech:2016xec}
\eqnRn{ConfBlockGhatRn}{
{1\over 4}\left(\beta^{(\Delta,\Delta)}\right)^2 \mathcal{G}_{\Delta}(u,v) = \int_{a \in \gamma_{12}} \int_{b \in \gamma_{34}} \hat{G}_{\Delta}(a,b)\,,
}
where $\hat{G}(a,b)$ is the unnormalized scalar bulk-to-bulk propagator in $\Rn$, and $\mathcal{G}_\Delta(u,v)$ is the scalar conformal block.

\section{Four-point contact and exchange diagrams}
\label{FOURPT}

In this section we introduce some $p$-adic propagator identities which greatly reduce the complexity of performing bulk integrations (more precisely, tree summations) encountered while evaluating various four-point amplitudes. The Archimedean analogs of these identities~\cite{Hijano:2015zsa} proved to be of great use in $\Rn$ for the evaluation of bulk integrals in the scalar four-point contact and exchange diagrams, and we show below how this carries over to $\Qq$.

\subsection{Two AdS propagator identities}
\label{IDENTITIES}
An identity which will be especially useful for decomposing bulk diagrams into geodesic bulk diagrams (and as a consequence of \eno{GeodesicConfWaveQq} and \eno{WDeltaAgain} into a conformal block decomposition) is
\eqnQq{KKaQq}{
 &\hat{K}_{\Delta_1}(b,x_1)  \hat{K}_{\Delta_2}(b,x_2) = a^{(\Delta_1,\Delta_2)} \cr
 & \times \sum_{a \in \gamma_{12}} \!\!\! \hat{K}_{\Delta_1}(a,x_1)  \hat{K}_{\Delta_2}(a,x_2) \hat{G}_{\Delta_1+\Delta_2}(a,b),
}
where $a^{(s,t)}$ is given by \eno{aDefQq}, and the bulk point $a$ is restricted to lie along $\gamma_{12}$, which is the unique bulk geodesic joining $x_1$ to $x_2$. This identity can be verified straightforwardly by explicit evaluation, but it's helpful to think about it geometrically as well. On the Bruhat--Tits tree, the paths from $x_1$ and $x_2$ to $b$ can be divided into two sub-paths each: the first which lies along the geodesic joining $x_1$ and $x_2$, and the second which lies off of that geodesic, and which is in fact common to both the paths from $x_1$ and $x_2$ to $b$. The r.h.s.~of \eno{KKaQq} can similarly be seen to decompose into subpaths: the bulk-to-boundary propagators are restricted to lie along the geodesic joining $x_1$ and $x_2$, while the bulk-to-bulk propagator travels partially along the geodesic, and partially along the common subpath mentioned above. The overall factor accounts for the over-counting of paths on the r.h.s. The corresponding identity in $\Rn$ is~\cite{Hijano:2015zsa}\footnote{There is an overall explicit factor of $2$ as compared with (4.1) of ~\cite{Hijano:2015zsa} due to a small difference in notation --- see footnote \ref{fn:notation}.}
\eqnRn{KKaRn}{
 &\hat{K}_{\Delta_1}(b,x_1) \hat{K}_{\Delta_2}(b,x_2) = 2\sum_{M=0}^\infty a_M^{(\Delta_1,\Delta_2)} \cr 
 & \times \int_{a\in \gamma_{12}} \!\!\!\!\! \hat{K}_{\Delta_1}(a,x_1)  \hat{K}_{\Delta_2}(a,x_2) \hat{G}_{\Delta_1 + \Delta_2 + 2M}(a,b),
}
where $a^{(s,t)}_M$ is given by \eno{aDefRn}. The crucial difference between \eno{KKaQq} and \eno{KKaRn} is the infinite sum over $M$ that has collapsed to the leading $M=0$ term in \eno{KKaQq}. The bulk-to-bulk propagators appearing in the identity represent a scalar of scaling dimension $\Delta$ with $\Delta = \Delta_1 + \Delta_2$ in $\Qq$, while in $\Rn$ one must perform a (weighted) sum over all scalars with $\Delta = \Delta_1 + \Delta_2 + 2M$ for all integral $M \geq 0$.

Another identity, which is extremely useful for replacing certain integrations over all of AdS with unintegrated expressions, takes the following form:
\eqnQq{IntGGQq}{
& \sum_{c \in T_{p^n}} \hat{G}_{\Delta_1}(a,c) \hat{G}_{\Delta_2}(b,c)\cr 
&\quad{} = {-\hat{G}_{\Delta_1}(a,b)/ \widetilde{c}_{\Delta_2}   + \hat{G}_{\Delta_2}(a,b)/\widetilde{c}_{\Delta_1}   \over m_{\Delta_1}^2- m_{\Delta_2}^2}\,,
}
where we remind the reader that $m_\Delta$ is the $p$-adic mass given by \eno{mDeltaQq}, and $\widetilde{c}_\Delta$ is given in \eno{ctDeltaQq}.
It is worth rewriting this identity in terms of the normalized bulk-to-bulk propagators~\cite{Gubser:2016guj},\footnote{Likewise in $\Rn$, $\widetilde{c}_\Delta$ given in \eno{ctDeltaRn} is the usual normalization constant of the bulk-to-bulk propagator. See, for example, equations (6.12) and (8.29) of \cite{DHoker:2002nbb}, and (133) of \cite{Gubser:2016guj}.}
 \eqnQq{Gnorm}{
 G_\Delta(a,b) = \widetilde{c}_{\Delta} \hat{G}_\Delta(a,b)\,,
 }
in which case it becomes
\eqnQq{IntGGnormQq}{
& \sum_{c \in T_{p^n}}\!\!\! {G}_{\Delta_1}(a,c) {G}_{\Delta_2}(b,c) = {{G}_{\Delta_2}(a,b)  - {G}_{\Delta_1}(a,b)   \over m_{\Delta_1}^2- m_{\Delta_2}^2}\,.
}
The corresponding identity satisfied by the unnormalized bulk-to-bulk propagators in $\Rn$ takes the form~\cite{Hijano:2015zsa}
\eqnRn{IntGGRn}{
\int_c \hat{G}_{\Delta_1}(a,c) \hat{G}_{\Delta_2}(b,c) = {\hat{G}_{\Delta_1}(a,b) - \hat{G}_{\Delta_2}(a,b) \over m_{\Delta_1}^2- m_{\Delta_2}^2}\,,
}
where the Archimedean mass is given by \eno{mDeltaRn}.
We list some more propagator identities in appendix \ref{PROPIDENTITIES}, which we will not have occasion to use in the present paper, but which may prove useful in evaluating higher-point correlators and bulk loop diagrams. 
 
  Curiously, despite having very different expressions, the masses in $\Qq$ and $\Rn$ subtract in a surprisingly similar manner. From the expression for the $p$-adic mass in \eno{mDeltaQq}, it follows
 \eqnQq{mSubtractQq}{
 m_{\Delta_A}^2 - m_{\Delta_B}^2 = {-p^{\Delta_B} \over \zeta(\Delta_B - \Delta_A) \zeta(\Delta_A + \Delta_B -n)}\,.
 }
This is to be compared with the Archimedean place, where
 \eqnRn{mSubtractRn}{
 m_{\Delta_A}^2-m_{\Delta_B}^2 = (\Delta_A - \Delta_B)(\Delta_A + \Delta_B - n)\,.
 }
This observation will prove useful later when we discuss and compare the the logarithmic singularity structure of the four-point function in $\Rn$ and $\Qq$.

\subsection{Four-point contact diagram, again}
\label{CONTACT}

We will now use \eno{KKaQq} and \eno{IntGGQq} to rederive \eno{DfnExpress}. Starting with \eno{Dfn} with the $x_i$ arranged in the $s$-channel configuration shown in figure \ref{figSchannel}, and applying identity \eno{KKaQq} to the pairs $\hat{K}_{\Delta_1}\hat{K}_{\Delta_2}$ and $\hat{K}_{\Delta_3}\hat{K}_{\Delta_4}$, we obtain
 \eqnQq{ContactStepOne}{
 &\mathcal{D} W_0 = a^{(\Delta_1, \Delta_2)} a^{(\Delta_3,\Delta_4)} \sum_{a \in T_{p^n}} \sum_{b_1 \in \gamma_{12}} \sum_{b_2 \in \gamma_{34}} \cr
 & \times \hat{K}_{\Delta_1}(b_1,x_1)\hat{K}_{\Delta_1}(b_1,x_2) \hat{G}_{\Delta_1+\Delta_2}(b_1,a) \cr
 & \times \hat{K}_{\Delta_3}(b_2,x_3)\hat{K}_{\Delta_4}(b_2,x_4) \hat{G}_{\Delta_3+\Delta_4}(b_2,a)\,.
 }
 While it may seem we have made our lives harder by introducing two additional summations over geodesics within the Bruhat--Tits tree, the effect is in fact the opposite: Thanks to the identities of sections \ref{GEODESICDIAG} and \ref{IDENTITIES}, we never have to explicitly evaluate {\it any} of these integrals. We first eliminate the sum over the bulk point $a$ by recognising it takes exactly the form of identity \eno{IntGGQq}. This results in two terms with the propagator content schematically of the form $\sim \sum_\gamma\sum_\gamma \hat{K}\hat{K} \hat{G}_\Delta \hat{K}\hat{K}$, with $\Delta= \Delta_1 + \Delta_2$ in one term, and $\Delta = \Delta_3 + \Delta_4$ in the other. This combination is exactly the same as \eno{GeoBulkDiag}, which is the definition of a geodesic bulk diagram. Substituting the geodesic bulk diagram with conformal partial waves using \eno{GeodesicConfWaveQq}, we arrive at 
\eqnQq{DfnConfBlockQq}{
\mathcal{D}(u) = P_1^{(12)} \mathcal{G}_{\Delta_A}(u) + P_1^{(34)} \mathcal{G}_{\Delta_B}(u)\,,
}
with the squared OPE coefficients
\begin{widetext}
\eqnQq{OPEsqQq}{
P_1^{(12)} &=   {-1\over \widetilde{c}_{\Delta_B}}\left(\beta^{(2\Delta_1,2\Delta_2)} a^{(\Delta_1,\Delta_2)}\right)  \left( \beta^{(\Delta_A+\Delta_{34},\Delta_A-\Delta_{34})}  {a^{(\Delta_3,\Delta_4)} \over m_{\Delta_A}^2 - m_{\Delta_B}^2}  \right)  \cr
P_1^{(34)} &=  {-1\over \widetilde{c}_{\Delta_A}} \left(\beta^{(2\Delta_3,2\Delta_4)} a^{(\Delta_3,\Delta_4)}\right)   \left( \beta^{(\Delta_B +\Delta_{12},\Delta_B - \Delta_{12})}  {a^{(\Delta_1,\Delta_2)} \over m_{\Delta_B}^2 - m_{\Delta_A}^2}  \right).
}
\end{widetext}
It is straightforward to check that \eno{DfnConfBlockQq} with the coefficients given in \eno{OPEsqQq} agrees precisely with \eno{DfnExpress}. That the expression for the $p$-adic four-point amplitude in \eno{DfnExpress} has an equivalent representation featuring $p$-adic mass singularities as shown in \eno{DfnConfBlockQq}-\eno{OPEsqQq} is highly non-trivial but physical. We comment more on this at the end of section \ref{TUEXCHANGE}.

A calculation in $\Rn$, similar to the one described for $\Qq$ using the propagator identities, results in an expression for the contact diagram similar to \eno{OPEsqQq}, but with an important difference. We quote the result computed in~\cite{Hijano:2015zsa}:
\eqnRn{DfnConfBlockRn}{
g(u,v) &= \sum_{M=0}^\infty P_1^{(12)}(M) \mathcal{G}_{\Delta_A +2M}(u,v) \cr 
 &+ \sum_{N=0}^{\infty} P_1^{(34)}(N) \mathcal{G}_{\Delta_B+2N}(u,v)\,,
}
where the squared OPE coefficients are given by
\begin{widetext}
\eqnRn{OPEsqRn}{
&P_1^{(12)}(M) = \left(\beta^{(2\Delta_1+2M,2\Delta_2+2M)} a_M^{(\Delta_1,\Delta_2)}\right) \left( \beta^{(\Delta_A+\Delta_{34}+2M,\Delta_A-\Delta_{34}+2M)} \sum_{N=0}^\infty {a_N^{(\Delta_3,\Delta_4)} \over m_{\Delta_A+2M}^2 - m_{\Delta_B+2N}^2}\right) \cr
&P_1^{(34)}(N) = \left(\beta^{(2\Delta_3+2N,2\Delta_4+2N)} a_N^{(\Delta_3,\Delta_4)}\right)  \left( \beta^{(\Delta_B+\Delta_{12}+2N,\Delta_B-\Delta_{12}+2N)} \sum_{M=0}^\infty {a_M^{(\Delta_1,\Delta_2)} \over m_{\Delta_B+2N}^2 - m_{\Delta_A+2M}^2}\right).
}
\end{widetext}
It can be seen in the conformal block decomposition of the four-point contact diagram in \eno{DfnConfBlockRn} that double-trace operators, schematically of the form $\mathcal{O}_i \partial^{2N} \mathcal{O}_j$ with scaling dimension $\Delta_i + \Delta_j + 2N$ at leading order, run in the intermediate channel. In $\Qq$, looking at \eno{DfnConfBlockQq}, we conclude that only double-trace operators {\it without} derivatives appear in the intermediate channel. (Essentially, in \eno{DfnConfBlockQq}-\eno{OPEsqQq}, the infinite sums over $M$ and $N$ in \eno{DfnConfBlockRn}-\eno{OPEsqRn} have collapsed to the $M=N=0$ term.) 
This is consistent with the general expectation that local derivatives of operators do not appear in $p$-adic CFTs.  This expectation stems in turn from the understanding that the ultrametric analog of a smooth function from reals to reals is a piecewise constant function from an ultrametric field to the reals.

\subsection{Exchange diagram in the direct channel}
\label{EXCHANGE}
In the rest of this section, we will use the previously stated propagator identities to evaluate four-point exchange diagrams. First we compute the diagram associated with the exchange of a scalar of dimension $\Delta$ in the $(12)(34)$ channel, which we will express in terms of a conformal block decomposition in the direct channel. Explicitly, we wish to evaluate
\eqnQq{ExchDiagS}{
\mathcal{D}_{\Delta}^{S}(x_i) &\equiv {1\over W_0}\! \sum_{a_1,a_2 \in T_{p^n}}\!\!\!\!\! \hat{K}_{\Delta_1}(a_1,x_1) \hat{K}_{\Delta_2}(a_1,x_2)\cr
& \times  \hat{G}_{\Delta}(a_1,a_2)  \hat{K}_{\Delta_3}(a_2,x_3) \hat{K}_{\Delta_4}(a_2,x_4).
}
Applying the propagator identity \eno{KKaQq} on the $\hat{K}_{\Delta_1} \hat{K}_{\Delta_2}$ and $\hat{K}_{\Delta_3} \hat{K}_{\Delta_4}$ legs leaves us with an expression involving the following double integration on the tree
\eqnQq{DoubleSum}{
\sum_{a_1, a_2 \in T_{p^n}} \!\!\!\!\! \hat{G}_{\Delta_A}(b_1,a_1) \hat{G}_{\Delta}(a_1,a_2)\hat{G}_{\Delta_B}(a_2,b_2),
}
where $b_1 \in \gamma_{12}$ and $b_2 \in \gamma_{34}$.
This can be immediately reduced to a combination of unintegrated bulk-to-bulk propagators by applying the identity \eno{IntGGQq} twice. Altogether, we wind up with
\begin{widetext}
\eqnQq{ExchSReduce}{
\mathcal{D}_{\Delta}^{S} &= {a^{(\Delta_1,\Delta_2)} a^{(\Delta_3,\Delta_4)} \over W_0 \widetilde{c}_\Delta \widetilde{c}_{\Delta_A} \widetilde{c}_{\Delta_B}} \sum_{b_1 \in \gamma_{12}} \sum_{b_2 \in \gamma_{34}} \! \Bigg[ \hat{K}_{\Delta_1}(b_1,x_1) \hat{K}_{\Delta_2}(b_1,x_2) \hat{K}_{\Delta_3}(b_2,x_3) \hat{K}_{\Delta_4}(b_2,x_4) \cr 
& \left. \times \! \left(\!{ \widetilde{c}_\Delta\hat{G}_\Delta(b_1,b_2)  \over (m_{\Delta}^2-m_{\Delta_A}^2)(m_{\Delta}^2-m_{\Delta_B}^2)} + {\widetilde{c}_{\Delta_A}\hat{G}_{\Delta_A}(b_1,b_2)  \over (m_{\Delta_A}^2-m_{\Delta}^2)(m_{\Delta_A}^2-m_{\Delta_B}^2)} + {\widetilde{c}_{\Delta_B}\hat{G}_{\Delta_B}(b_1,b_2) \over (m_{\Delta_B}^2-m_{\Delta}^2)(m_{\Delta_B}^2-m_{\Delta_A}^2)} \! \right) \! \right]\!.
}
Recognizing the integral over points $b_1, b_2$ as the geodesic bulk diagram defined in \eno{GeoBulkDiag}, and using \eno{GeodesicConfWaveQq} and \eno{WDeltaAgain} to express in terms of conformal blocks, we obtain
\eqnQq{ExchSFinal}{
\mathcal{D}_{\Delta}^{S}(u) &= 
 C_{12 \Delta} C_{34 \Delta}\: \mathcal{G}_{\Delta}(u) + 
P_1^{(12)} \mathcal{G}_{\Delta_A}(u) + P_1^{(34)} \mathcal{G}_{\Delta_B}(u)
}
where
\eqnQq{OPEsqExch}{
 C_{12 \Delta} C_{34 \Delta} &= {1 \over \widetilde{c}_{\Delta_A}\widetilde{c}_{\Delta_B}} \left(\beta^{(\Delta+\Delta_{12},\Delta-\Delta_{12})} {a^{(\Delta_1,\Delta_2)} \over m_{\Delta}^2 - m_{\Delta_A}^2}  \right) \left(\beta^{(\Delta+\Delta_{34},\Delta-\Delta_{34})} {a^{(\Delta_3,\Delta_4)} \over m_{\Delta}^2 - m_{\Delta_B}^2}  \right) \cr
P_1^{(12)} &=  {1 \over \widetilde{c}_{\Delta}\widetilde{c}_{\Delta_B}}  \left(\beta^{(2\Delta_1,2\Delta_2)} {a^{(\Delta_1,\Delta_2)} \over m_{\Delta_A}^2 - m_{\Delta}^2}  \right) \left(\beta^{(\Delta_A+\Delta_{34},\Delta_A-\Delta_{34})} {a^{(\Delta_3,\Delta_4)} \over m_{\Delta_A}^2 - m_{\Delta_B}^2}  \right) \cr
P_1^{(34)} &=  {1 \over \widetilde{c}_{\Delta}\widetilde{c}_{\Delta_A}} \left(\beta^{(2\Delta_3,2\Delta_4)} {a^{(\Delta_3,\Delta_4)} \over m_{\Delta_B}^2 - m_{\Delta}^2}  \right) \left(\beta^{(\Delta_B+\Delta_{12},\Delta_B-\Delta_{12})}  {a^{(\Delta_1,\Delta_2)} \over m_{\Delta_B}^2 - m_{\Delta_A}^2}  \right).
 }
\end{widetext}
This is the $p$-adic analog of equations (4.16)--(4.17) in \cite{Hijano:2015zsa}. The similarities between the $p$-adic and Archimedean OPE coefficients (squared) are remarkable. In the conformal block decomposition in the direct channel, in addition to the double-trace exchanges (albeit without derivatives just like in the case of the contact diagram), we find  as expected, a term representing the single-trace exchange of a scalar of dimension $\Delta$.

\subsection{Exchange diagrams in the crossed channel}
\label{TUEXCHANGE}
 \begin{figure}[t]
  \centerline{\includegraphics[width=2.3in]{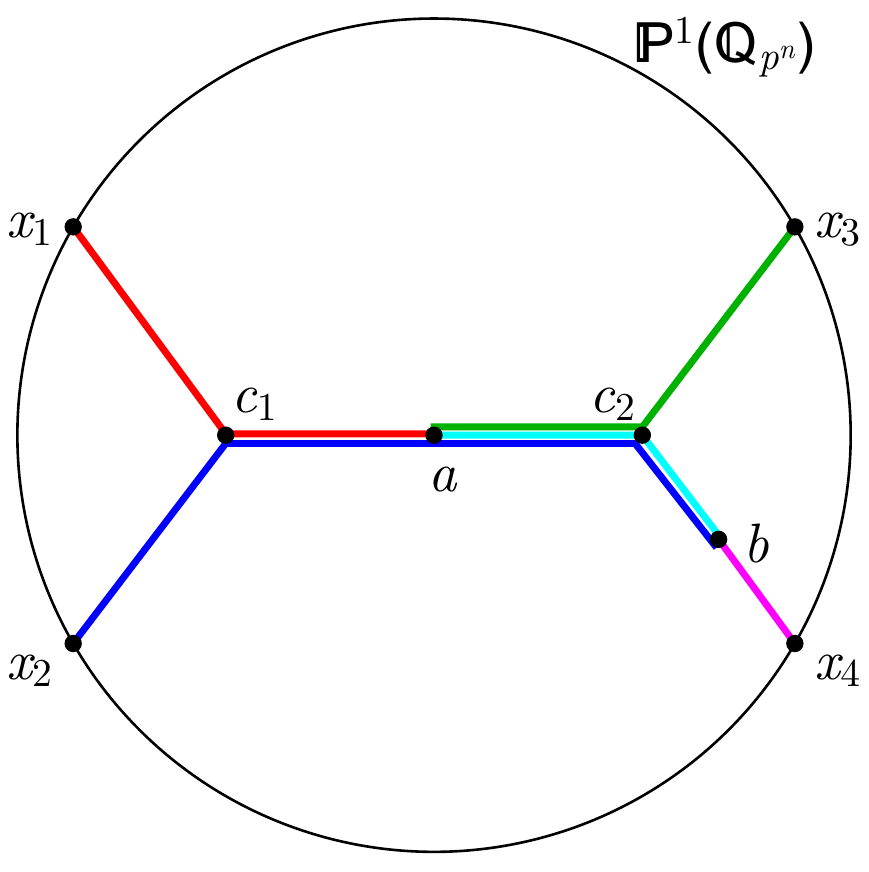}}
  \caption{(Color online.) Geodesic subway diagram ${\cal W}_\Delta^T$, with exchange of a scalar in the $(13)(24)$ channel. The bulk point $a$ runs along the geodesic joining $x_1$ with $x_3$, and $b$ runs along the geodesic joining $x_2$ with $x_4$, while a scalar of dimension $\Delta$ is exchanged between $a$ and $b$. Colors differentiate the
individual propagators found in ${\cal W}_\Delta^T$ in \eno{GeodesicbulkDiagCrossed}.}\label{figGeodesicT}
 \end{figure}
In section \ref{GEODESICDIAG}, we computed the geodesic bulk diagram for boundary points $x_i$ in an `$s$-channel' configuration (i.e. $u<1$) with a scalar of dimension $\Delta$ exchanged in the $(12)(34)$ channel (see figure \ref{figGeodesic}). Equation \eno{GeodesicConfWaveQq} then states that up to an overall factor, the geodesic bulk diagram is simply the conformal partial wave $W_{\Delta}(x_i)$. Let's now consider two closely related geodesic bulk diagrams which will prove useful for computing exchange diagrams in the crossed channels. In these geodesic diagrams the boundary points $x_i$ remain in the `$s$-channel' configuration shown in figure \ref{figSchannel} with $u<1$, but a scalar of dimension $\Delta$ is exchanged in the $(13)(24)$ channel, or the $(14)(23)$ channel. Explicitly, we define these geodesic bulk diagrams to be (see figure \ref{figGeodesicT})
\eqnQq{GeodesicbulkDiagCrossed}{
\mathcal{W}_{\Delta}^T(x_i) &\equiv \sum_{\substack{a\in \gamma_{13} \\ b\in \gamma_{24}}} \hat{K}_{\Delta_1}(x_1,a)\hat{K}_{\Delta_3}(x_3,a) \hat{G}_{\Delta}(a,b) \cr  & \times \hat{K}_{\Delta_2}(x_2,b) \hat{K}_{\Delta_4}(x_4,b) \cr
\mathcal{W}_{\Delta}^U(x_i) &\equiv \sum_{\substack{a\in \gamma_{14} \\ b\in \gamma_{23}}} \hat{K}_{\Delta_1}(x_1,a)\hat{K}_{\Delta_4}(x_4,a) \hat{G}_{\Delta}(a,b) \cr
& \times \hat{K}_{\Delta_2}(x_2,b) \hat{K}_{\Delta_3}(x_3,b)\,.
} 
A direct computation on the Bruhat--Tits tree, detailed in appendix \ref{MOREDETAILS}, reveals the following decomposition of a geodesic bulk diagram in the crossed-channel,
\eqnQq{GeodesicConfWaveT}{
\mathcal{W}_{\Delta}^T &= \beta^{(\Delta+\Delta_{13},\Delta-\Delta_{13})} \beta^{(-\Delta_{13}-\Delta_{24},\Delta+\Delta_{24})} W_{\Delta_A} \cr 
 &\quad + \beta^{(\Delta+\Delta_{13},\Delta-\Delta_{13})} \beta^{(\Delta_{13}+\Delta_{24},\Delta - \Delta_{24})}  W_{\Delta_B} \cr
& = \beta^{(\Delta +\Delta_{24},\Delta-\Delta_{24})} \beta^{(-\Delta_{13}-\Delta_{24},\Delta+\Delta_{13})}  W_{\Delta_A} \cr
&\quad + \beta^{(\Delta +\Delta_{24},\Delta-\Delta_{24})} \beta^{(\Delta_{13}+\Delta_{24},\Delta - \Delta_{13})}  W_{\Delta_B}.
}
The corresponding identity for a geodesic diagram with exchange in the $(14)(23)$ channel, $\mathcal{W}_\Delta^U$, is obtained simply by switching $\Delta_3 \leftrightarrow \Delta_4$ in \eno{GeodesicConfWaveT}.

  A non-trivial consistency check on \eno{GeodesicConfWaveT} can be obtained by starting from the defining expression \eno{Dfn} for the contact diagram, then using \eno{KKaQq} on the $\hat{K}_{\Delta_1}\hat{K}_{\Delta_3}$ and $\hat{K}_{\Delta_2}\hat{K}_{\Delta_4}$ legs, then applying \eno{IntGGQq} once, and finally employing \eno{GeodesicConfWaveT} to obtain \eno{DfnExpress}.  This slightly round-about method is easily seen to agree with the simpler calculation outlined in section \ref{CONTACT}.

We can now use the identity \eno{GeodesicConfWaveT} to evaluate the four-point exchange diagram where a scalar of dimension $\Delta$ is exchanged in the $(13)(24)$ channel (we remind the readers that the boundary points $x_i$ will always be in the `$s$-channel' configuration shown in figure \ref{figGeodesicT}, i.e.~$u<1$):
\eqnQq{ExchangeDiagCrossed}{
\mathcal{D}_{\Delta}^{T}(x_i) &\equiv {1\over W_0} \! \sum_{a_1,a_2 \in T_{p^n}} \!\!\! \hat{K}_{\Delta_1}(a_1,x_1) \hat{K}_{\Delta_3}(a_1,x_3) \cr 
& \times \hat{G}_{\Delta}(a_1,a_2)  \hat{K}_{\Delta_2}(a_2,x_2) \hat{K}_{\Delta_4}(a_2,x_4)\,.
}
We will express the final result in terms of a conformal block decomposition in the crossed channel.
The computation proceeds along lines similar to the one  for $\mathcal{D}_\Delta^S$ sketched in section \ref{EXCHANGE}, and is described in appendix \ref{MOREDETAILS}. The final result is 
\begin{widetext}
\eqnQq{ExchTFinal}{
\mathcal{D}_{\Delta}^{T}(u) = P_1^{(12)} \mathcal{G}_{\Delta_A}(u) + P_1^{(34)} \mathcal{G}_{\Delta_B}(u),
}
where 
\eqnQq{OPEsqExchT12}{
 P_1^{(12)} & = {1 \over \widetilde{c}_{\Delta_C} \widetilde{c}_{\Delta_D}} \left(\beta^{(\Delta+\Delta_{13},\Delta-\Delta_{13})} {a^{(\Delta_1,\Delta_3)} \over m_{\Delta}^2 - m_{\Delta_C}^2}  \right) \left(\beta^{(-\Delta_{13}-\Delta_{24}, \Delta+\Delta_{24})} {a^{(\Delta_2,\Delta_4)} \over m_{\Delta}^2 - m_{\Delta_D}^2}  \right) \cr
  & + {1 \over \widetilde{c}_{\Delta} \widetilde{c}_{\Delta_D}} \left(\beta^{(2\Delta_1,2\Delta_3)} {a^{(\Delta_1,\Delta_3)} \over m_{\Delta_C}^2 - m_{\Delta}^2}  \right) \left(\beta^{(-\Delta_{13}-\Delta_{24},\Delta_C+\Delta_{24})} {a^{(\Delta_2,\Delta_4)} \over m_{\Delta_C}^2 - m_{\Delta_D}^2}  \right) \cr
  & + {1 \over \widetilde{c}_{\Delta} \widetilde{c}_{\Delta_C}}\left(\beta^{(2\Delta_2,2\Delta_4)} {a^{(\Delta_2,\Delta_4)} \over m_{\Delta_D}^2 - m_{\Delta}^2}  \right) \left(\beta^{(-\Delta_{13}-\Delta_{24},\Delta_D+\Delta_{13})} {a^{(\Delta_1,\Delta_3)} \over m_{\Delta_D}^2 - m_{\Delta_C}^2}  \right),
}
and
\eqnQq{OPEsqExchT34}{
 P_1^{(34)} & = {1 \over \widetilde{c}_{\Delta_C} \widetilde{c}_{\Delta_D}}\left(\beta^{(\Delta+\Delta_{13},\Delta-\Delta_{13})} {a^{(\Delta_1,\Delta_3)} \over m_{\Delta}^2 - m_{\Delta_C}^2} \right) \left(\beta^{(\Delta_{13}+\Delta_{24},\Delta-\Delta_{24})} {a^{(\Delta_2,\Delta_4)} \over m_{\Delta}^2 - m_{\Delta_D}^2}  \right) \cr
  & + {1 \over \widetilde{c}_{\Delta} \widetilde{c}_{\Delta_D}} \left(\beta^{(2\Delta_1,2\Delta_3)} {a^{(\Delta_1,\Delta_3)} \over m_{\Delta_C}^2 - m_{\Delta}^2}  \right) \left(\beta^{(\Delta_{13}+\Delta_{24},\Delta_C-\Delta_{24})} {a^{(\Delta_2,\Delta_4)} \over m_{\Delta_C}^2 - m_{\Delta_D}^2}  \right) \cr
  & + {1 \over \widetilde{c}_{\Delta} \widetilde{c}_{\Delta_C}} \left(\beta^{(2\Delta_2,2\Delta_4)} {a^{(\Delta_2,\Delta_4)} \over m_{\Delta_D}^2 - m_{\Delta}^2} \right) \left(\beta^{(\Delta_{13}+\Delta_{24},\Delta_D-\Delta_{13})} {a^{(\Delta_1,\Delta_3)} \over 
m_{\Delta_D}^2 - m_{\Delta_C}^2
}  \right) .
 }
\end{widetext}
Here we have defined
\eqn{DeltaCD}{
\Delta_C = \Delta_1 + \Delta_3 \qquad \Delta_D = \Delta_2 + \Delta_4\,.
}
 From a diagrammatic point of view in the bulk it appears that $\mathcal{G}_\delta(u)$ for various other values of $\delta$, like $\Delta_1 + \Delta_4 +\Delta$ or $\Delta_2 + \Delta_3 +\Delta$, might appear in the intermediate steps while computing \eno{ExchangeDiagCrossed}. But miraculously these contributions wind up canceling in the final result,
and as expected for exchange diagrams expressed in the conformal block decomposition in a crossed-channel, only double-trace exchanges appear. (See appendix \ref{CROSSING} for an explanation of this point and related comments on crossing symmetry.) The exchange diagram $\mathcal{D}_{\Delta}^{U}$, where a scalar of dimension $\Delta$ is exchanged in the $(14)(23)$ channel is obtained from \eno{ExchTFinal} simply by switching $\Delta_3 \leftrightarrow \Delta_4$.

It is worth pointing out that in $\Rn$, anomalous dimensions appear in the tree-level four-point contact amplitude in the form of logarithmic singularities when the integrality condition, $\Delta_A - \Delta_B \in 2\mathbb{Z}$ is met, or equivalently when the algebraic condition $m_{\Delta_A+2M}^2 = m_{\Delta_B+2N}^2$ is satisfied in \eno{OPEsqRn} for integral $M,N \geq 0$~\cite{Hijano:2015zsa}.  Instead in $\Qq$, we find logarithmic singularities arise only when $\Delta_A - \Delta_B = 0$, or equivalently when $m_{\Delta_A}^2 = m_{\Delta_B}^2$.\footnote{If alternate quantization is allowed, it is clear from \eno{mSubtractQq}-\eno{mSubtractRn} that the condition $m_{\Delta_A+2M}^2 = m_{\Delta_B+2N}^2$ has in addition to $\Delta_A - \Delta_B \in 2\mathbb{Z}$, a second solution, $\Delta_A + \Delta_B - n = 2\ell$ where $\ell$ is a non-positive integer. (In the $p$-adics, $M=N=0$, so the conditions are more restrictive: $\Delta_A - \Delta_B = 0$ or $\Delta_A + \Delta_B - n=0$.)  For $n > 4$, the second solution is disallowed since it violates the unitarity bound, which restricts $\Delta_A, \Delta_B \geq n-2$. For the second solution to exist in $n=4$, $\Delta_A$ and $\Delta_B$ must saturate the unitarity bound, so $\Delta_A = \Delta_B = 2$ and $\Delta_A - \Delta_B \in 2\mathbb{Z}$ is satisfied. However, for $n \leq 3$ there exist pairs of scaling dimensions satisfying the unitarity bound, such that  $\Delta_A + \Delta_B = n$ but $\Delta_A - \Delta_B \notin 2\mathbb{Z}$. Such exceptional choices would seem to hint at the appearance of a new kind of logarithmic singularity with an origin different from the usual integrality condition $\Delta_A - \Delta_B \in 2\mathbb{Z}$. In the $p$-adics, the convergence conditions listed in footnote \ref{fn:convergence} portend the appearance of severe singularities in the four-point contact amplitude if $\Delta_A + \Delta_B \leq n$. So for example, \eno{DfnExpress} or equivalently \eno{DfnConfBlockQq}-\eno{OPEsqQq} is altogether not to be trusted when $\Delta_A + \Delta_B - n \leq 0$, and we cannot reasonably inquire about singularities at special values.  Might a similar argument in $\Rn$ prevent the appearance of these exceptional singularities?} 
 This is intriguingly reminiscent of the existence of an infinite sequence of poles in the anomalous dimension of composite operators in CFTs in $\Rn$ as opposed to just one pole in $p$-adic CFTs, at least in the context of the $O(N)$ model~\cite{Gubser:2017vgc}. Analogously for the exchange diagrams, logarithmic singularities arise in the exchange amplitude \eno{ExchSFinal}-\eno{OPEsqExch} when any of $m^2_\Delta, m^2_{\Delta_A}, m^2_{\Delta_B}$ coincide, and in \eno{ExchTFinal}-\eno{OPEsqExchT34} simply when any of $m^2_\Delta, m^2_{\Delta_C}, m^2_{\Delta_D}$ coincide.

\section{Towards a bulk dual of free field theory}
\label{FREECFT}

 So far in this paper, we have presented, in the context of $p$-adic AdS/CFT, the holographic computation of the four-point contact and exchange diagrams for scalar composite operators of general dimensions. 
In this section we would like to construct a minimal bulk theory that reproduces the correlators of a free $p$-adic field theory, featuring an operator ${\cal O}$ of dimension $\Delta$. 

In the minimal construction, we would include only one bulk field, namely a scalar $\phi$ with mass squared $m^2_\Delta$ as defined in \eno{mDeltaQq}, and with only cubic interactions.  It turns out this is not enough to give a four-point function that agrees with a free $p$-adic field theory.  As a next-to-minimal construction, we could consider adding quartic interactions for $\phi$.  This is still not enough to reproduce the four-point function of a free boundary theory.  As we will explain, a strategy which {\it does} work (at least as far as the four-point function) is to include also quartic interactions for $\phi$ which act across a link: that is, nearest neighbor interactions. An efficient way to package all the constructions we have in mind is to introduce an additional bulk scalar $\tilde\phi$ whose scaling dimension $\widetilde\Delta$ we will eventually take to be large, and to allow only cubic on-site interactions of the form $\phi^3$ and $\tilde\phi \phi^2$. The action for such a $\phi$-$\tilde\phi$ theory takes the form
 \eqnQq{ActionPhiPhit}{
 S[\phi,\tilde\phi] &=  \sum_{\langle ab \rangle} {1 \over 2} (\phi_a - \phi_b)^2 + \sum_{\langle ab \rangle} {1 \over 2} (\tilde\phi_a - \tilde\phi_b)^2 \cr 
 &+ \sum_{a \in T_{p^n}} \!\left( {1\over 2} m_{\Delta}^2 \phi_a^2 + {1\over 2} m_{\widetilde\Delta}^2 \tilde\phi_a^2  \right. \cr
 & \qquad \left. + {g_3 \over 3!} \phi_a^3  + {\widetilde g_3 \over 2} \phi_a^2 \tilde\phi_a \right).
 }
  In the strict limit of large $m_{\widetilde \Delta}$, any diagram where $\tilde\phi$ propagates even a single step becomes negligible.

Specialising in \eno{ExchSFinal} and \eno{ExchTFinal} to the case of all four $\Delta_i = {\Delta}$ with the dimension of the exchanged scalar relabelled $\widetilde\Delta$, we get, after summing up the exchange contributions from all channels,
 \eqnQq{DExchangeDeltat}{
 \mathcal{D}_{\widetilde\Delta}^{(\rm exchange)} &\equiv \mathcal{D}_{\widetilde\Delta}^S + \mathcal{D}_{\widetilde\Delta}^T + \mathcal{D}_{\widetilde\Delta}^U \cr  &= {F}_1 u^{\widetilde\Delta} + {F}_2 u^{2\Delta} + {F}_3 u^{2\Delta} {\log u \over \log p}\,,
 }
where the constants ${F}_i$ depend on $n, p, \Delta$ and $\widetilde\Delta$ but not on $u$.  The $F_i$ can be evaluated immediately from \eno{ExchSFinal}-\eno{OPEsqExch} and \eno{ExchTFinal}-\eno{OPEsqExchT34}, but the explicit form is complicated enough as to be unenlightening at this stage.  In \eno{DExchangeDeltat} we are not assuming large $m_{\widetilde{\Delta}}$. 
The first term in \eno{DExchangeDeltat} indicates, heuristically, that an operator of dimension $\widetilde\Delta$ can participate in the connected four-point function of an operator ${\cal O}$ of dimension $\Delta$, while the second two terms are evidence that one or more operators with dimensions close to $2\Delta$ participate.

 We now show explicitly that integrating out $\tilde\phi$ results in the contact diagram.  If we set
 \eqnQq{yDef}{
  y = p^{-\widetilde\Delta} \,,
 }
then we find
 \eqnQq{FcoefsOrder}{
  F_1 = O(y^2) \qquad F_2 = O(1) \qquad {F}_3 = O(1) \,.
 }
Setting $y=0$ is the same as $\widetilde\Delta \to \infty$, and because of \eno{FcoefsOrder} it gives a finite limit:
 \eqnQq{DcontactLimit}{
 &{\cal D}^{(\rm contact)} \equiv \lim_{\widetilde\Delta \to \infty} {\cal D}^{(\rm exchange)}_{\widetilde\Delta} 
    = {\cal D}^{(\rm exchange)}_{\widetilde\Delta} \Big|_{y=0} \cr 
    &= 
      {3 (1 + 4 p^{2\Delta} + p^{4\Delta} + d (-1 + p^{4\Delta})) \over 
        -p^n + p^{4\Delta}} u^{2\Delta}\,,
 }
where $d \equiv -\log u / \log p$.  A useful check is to note that ${\cal D}^{(\rm contact)} = 3{\cal D}|_{\Delta_i = \Delta}$, where ${\cal D}$ is the four-point contact amplitude given in \eno{DfnConfBlockQq}.\footnote{Moreover, ${\cal D}^{(\rm contact)} = 3D_p/W_0$, where $D_p$ is the four-point contact amplitude evaluated in eq.~(122) of \cite{Gubser:2016guj}.}

If instead of setting $y=0$ we pick out the $O(y)$ term of ${\cal D}^{(\rm exchange)}_{\widetilde\Delta}$ at large $\widetilde\Delta$ (meaning small $y$), it means we are focusing on nearest neighbor interactions, i.e. an interaction which takes place when two bulk points are precisely one step apart.  Thus we define
 \eqnQq{DnearestLimit}{
 & {\cal D}^{(\rm nearest)} \equiv {d {\cal D}^{(\rm exchange)}_{\widetilde\Delta} \over dy}
    \Bigg|_{y=0} \cr 
    &= (p^{2\Delta}+p^{n-2\Delta}){\cal D}^{(\rm contact)} - u^{2\Delta}\left( {3 \over \widetilde{c}_{2\Delta}} + {2d \over \widetilde{c}_{\Delta}^2}\right)\!,
 }
where $\widetilde{c}_\Delta$ is given by \eno{ctDeltaQq}.
 
To gain more intuition on the nearest neighbor interaction, it helps to arrive at \eno{DnearestLimit} from a different starting point. Define the nearest neighbor exchange amplitudes,
  \eqn{FourPtNN}{
  \mathcal{F}_{S} &\equiv \!{1 \over W_0}\!\! \sum_{a \in T_{p^n}}\! \sum_{b \sim a}\! \hat{K}_{\Delta}(x_1,a) \hat{K}_{\Delta}(x_2,a) \hat{K}_{\Delta}(x_3,b) \hat{K}_{\Delta}(x_4,b) \cr
    \mathcal{F}_{T} &\equiv \!{1 \over W_0}\!\! \sum_{a \in T_{p^n}}\! \sum_{b \sim a}\! \hat{K}_{\Delta}(x_1,a) \hat{K}_{\Delta}(x_2,b) \hat{K}_{\Delta}(x_3,a) \hat{K}_{\Delta}(x_4,b) \cr
      \mathcal{F}_{U} &\equiv \!{1 \over W_0}\!\! \sum_{a \in T_{p^n}}\! \sum_{b \sim a}\! \hat{K}_{\Delta}(x_1,a) \hat{K}_{\Delta}(x_2,b) \hat{K}_{\Delta}(x_3,b) \hat{K}_{\Delta}(x_4,a).
  }
  Here $\sum_{b \sim a}$ represents summing over all nearest neighbors of $a$, with $a$ held fixed. Thus, nearest neighbor interactions are manifest in the amplitudes \eno{FourPtNN}. In fact, one can easily check that $\mathcal{D}^{(\rm nearest)} = \mathcal{F}_S + {\cal F}_T + {\cal F}_U$. This was expected precisely because $\mathcal{D}^{(\rm nearest)}$ represents a nearest neighbor interaction where two pairs of propagators meet one edge apart, and summing the amplitudes in \eno{FourPtNN} accounts for all the possible ways that may happen. We briefly discuss the connection between nearest neighbor interactions and derivative couplings in appendix \ref{CUBICDERIV}.
 
 Now we observe that including both contact and nearest neighbor interactions allows us to form a four-point function where we can control $F_2$ and ${F}_3$, where $F_1, F_2$ and $F_3$ are the coefficients of $u^{\Delta}, u^{2\Delta}$ and $u^{2\Delta} \log_p u$, respectively.  In particular, we can cancel off ${F}_3$ and control the ratio $F_2/F_1$.  The interesting case to consider is the amplitude
 \eqnQq{Dcombined}{
  {\cal D}^{(\rm combined)} &\equiv  - 
    {(1+p^\Delta)^2 (-p^n+p^{4\Delta}) \over 2 (p^n-p^{3\Delta})^2} {\cal D}^{(\rm contact)} \cr 
    & +{\cal D}^{(\rm exchange)}_\Delta - {p^{2\Delta} (-1+p^{2\Delta}) \over 2 (p^n - p^{3\Delta})^2} {\cal D}^{(\rm nearest)}  \cr
   &= {f_{\Delta\Delta\Delta}^2 \over 2} \left( 2 u^\Delta + u^{2\Delta} \right),
 }
 where the structure constant $f_{\Delta\Delta\Delta}$ is given by \eno{fijk}.
The coefficients of ${\cal D}^{(\rm contact)}$ and ${\cal D}^{(\rm nearest)}$ in \eno{Dcombined} were chosen carefully so that the $u$-dependence of ${\cal D}^{(\rm combined)}$ would be exactly the $2 u^\Delta + u^{2\Delta}$ behavior expected in case the operator ${\cal O} = \vec\Phi^2$ where $\vec\Phi$ is a free field on the boundary.\footnote{To see this, note that
$$ u^\Delta W_{0}(x_i) = {1 \over |x_{12} x_{24} x_{34} x_{13} |^\Delta}
     = {1 \over |x_{12} x_{23} x_{34} x_{41}|^{\Delta}} $$
$$ u^{2\Delta} W_{0}(x_i) = {1 \over |x_{13} x_{24}|^{2\Delta}}
     = {1 \over |x_{13} x_{23} x_{24} x_{14}|^\Delta}\,,
$$
where in the second and fourth equalities we used $|x_{14} x_{23}| = |x_{13} x_{24}|$  (which is simply a rephrasing of $v = 1$). These account for the three possible Wick contractions we expect to see in the connected four-point function of a free theory.}
   
Having found in \eno{Dcombined} a combination of bulk amplitudes suggestive of a free field dual, we should next inquire what bulk action leads to \eno{Dcombined}.  It is awkward to use \eno{ActionPhiPhit} because to get ${\cal D}^{(\rm nearest)}$ from it we require the derivative operation \eno{DnearestLimit} in the $\widetilde{\Delta} \to \infty$ limit.  Let us therefore start instead from the action
\eqnQq{Snearest}{
S[\phi] &=  \sum_{\langle ab \rangle} {1 \over 2} (\phi_a - \phi_b)^2 \cr &+ \sum_{a \in T_{p^n}} \!\left( {1\over 2} m_{\Delta}^2 \phi_a^2 + {g_3 \over 3!} \phi_a^3 + {g_4 \over 4!} \phi_a^4 \right) \cr &+
 {\widetilde g_4 \over 8} \sum_{a \in T_{p^n}}\sum_{b \sim a} \phi_a^2 \phi_b^2\,,
}
which is precisely \eno{Ssingle} augmented by a nearest neighbor interaction.  Straightforward diagrammatic considerations lead us from \eno{Snearest} to
  \eqnQq{FourPtFree}{
 {1 \over W_0} &\langle {\cal O}(x_1){\cal O}(x_2){\cal O}(x_3){\cal O}(x_4)\rangle  \cr
 & =  -g_4 \widetilde{c}_{\Delta}^2 {{\cal D}^{(\rm contact)} \over 3}+{g_3^2} \widetilde{c}_{\Delta}^3 {\cal D}^{(\rm exchange)} \cr
 & \quad - \widetilde g_4 \widetilde{c}_{\Delta}^2 {\cal D}^{(\rm nearest)}\,.
 }
The factors of $\widetilde{c}_{\Delta}$ arise because each external leg picks up a factor of $\sqrt{\widetilde{c}_{\Delta}}$ (c.f.~the discussion around \eno{ThreePtRn}-\eno{ctDeltaQq}) and an extra factor of $\widetilde{c}_{\Delta}$ comes from the bulk-to-bulk propagator, as in \eno{Gnorm}.  The factor of $1/3$ in the first term comes from the relation ${\cal D}^{(\rm contact)} = 3 {\cal D}|_{\Delta_i=\Delta}$.
Choosing the couplings to be
 \eqnQq{CouplingsFree}{
 g_4 &= {3g_3^2 \over 2} { m_{3\Delta}^2 - m_{\Delta}^2 \over \left(m_{2\Delta}^2 - m_{\Delta}^2\right)^2} \cr 
 \widetilde g_4 &= {g_3^2 \over 2}{\left(\beta^{(\Delta,-4\Delta)} \beta^{(\Delta,3\Delta)}\right)^{-1} \over \left(m_{2\Delta}^2 - m_{\Delta}^2\right)^2} \,,
 }
we arrive at the connected four-point function of a free theory,
 \eqnQq{FourPtFreeFinal}{
  \langle {\cal O}(x_1)&{\cal O}(x_2){\cal O}(x_3){\cal O}(x_4)\rangle  \cr
  & ={g_3^2 \over 2}\widetilde{c}_{\Delta}^3  f_{\Delta\Delta\Delta}^2  (2u^{\Delta}+u^{2\Delta})W_0\,.
  }

Now, from the boundary perspective, ${\cal O} = \vec{\Phi}^2$ where $\vec{\Phi}$ is a free-field on the boundary, with the propagator
 \eqnQq{PhiProp}{
 \langle \Phi^I(x_1) \Phi^J(x_2) \rangle = {C \delta^{IJ}  \over |x_{12}|^{2\Delta_\Phi}}\,,
 }
 for some constant $C$, and $I,J = 1,\ldots, N$. We set $C = 1/\sqrt{2N}$.
It then follows that the two- and three-point functions of the composite operator ${\cal O}$ are (up to contact terms)
 \eqnQq{OObdy}{
 \langle {\cal O}(x_1) {\cal O}(x_2) \rangle &= \langle \Phi^I(x_1) \Phi^I(x_1) \Phi^J(x_2)\Phi^J(x_2) \rangle \cr
 &= {1 \over |x_{12}|^{4\Delta_{\Phi}}}
 }
 and
 \eqnQq{OOObdy}{
 \langle {\cal O}(x_1)  {\cal O}(x_2) {\cal O}(x_3) \rangle = {2\sqrt{2} \over \sqrt{N}}{ 1\over |x_{12}x_{23}x_{31}|^{2\Delta_{\Phi}}}\,,
 }
where $2\sqrt{2}/\sqrt{N} = 8C^3 N$, with the factor of $8$ coming from the possible Wick contractions. 
 On the other hand, the holographically obtained two- and three-point functions are\footnote{\label{fn:translate}The three-point function in \eno{TwoThreePtFree} comes from \eno{ThreePtQq}.  The calculation of the two-point function is slightly subtle and is discussed in detail in \cite{Gubser:2016guj}.  To translate from \cite{Gubser:2016guj} to our current conventions, set $\eta_p=1$ and ${\cal O}_{\rm here} = {\sqrt{\widetilde{c}_{\Delta}} / c_\Delta} {\cal O}_{\rm there}$ where $c_\Delta$, $\widetilde{c}_\Delta$ are given in \eno{cDelta} and \eno{ctDeltaQq} respectively.  This rescaling leads directly to $\langle {\cal O}_{\rm here}(x_1) {\cal O}_{\rm here}(x_2) \rangle = 1/|x_{12}|^{2\Delta}$.}
 \eqnQq{TwoThreePtFree}{
 \langle {\cal O}(x_1){\cal O}(x_2) \rangle &= {1 \over |x_{12}|^{2\Delta}} \cr
\langle {\cal O}(x_1){\cal O}(x_2){\cal O}(x_3) \rangle &= {-g_3 \widetilde{c}_{\Delta}^{3/2} f_{\Delta\Delta\Delta} \over |x_{12}x_{23}x_{31}|^\Delta}\,.
 }
 Equating \eno{OObdy}-\eno{OOObdy} with the holographic correlators in \eno{TwoThreePtFree} we conclude $\Delta = 2\Delta_{\Phi}$, with
 \eqnQq{g3Match}{
 g_3 = {-1 \over \sqrt{N}} {2 \sqrt{2} \over \widetilde{c}_\Delta^{3/2} f_{\Delta\Delta\Delta}}\,.
 }
In a free $p$-adic CFT, $\Delta_\Phi = (n-s)/2$ where $s$ is a (continuous) free parameter and is usually restricted to be in the range $n/2 < s < n$~\cite{Gubser:2017vgc}.  Using \eno{g3Match}, referring to \eno{fijk} for the explicit form of $f_{\Delta\Delta\Delta}$, and setting $\Delta=2\Delta_\Phi$, we see that $g_3$ vanishes at $n=3s/2$, while $g_4$ and $\widetilde g_4$ in \eno{CouplingsFree} stay finite and non-vanishing there. It is interesting to compare this with the situation in the Archimedean case, where we usually set $s=2$ and the cubic scalar coupling vanishes at $n=3$~\cite{Petkou:1994ad,Sezgin:2003pt}.\footnote{Curiously, the $p$-adic couplings $g_3, g_4$ and $\widetilde g_4$ vanish simultaneously for $n=s$. Could this be related to higher spin theories at $n=s=2$ (i.e.~AdS$_3$) in the Archimedean case, which are known to have special properties \cite{Vasiliev:1995dn, Gaberdiel:2010pz, Gaberdiel:2012uj}?}

 Continuing on to the connected free-field four-point function, up to contact terms, it is
 \eqnQq{OOOObdy}{
 & \langle {\cal O}(x_1)  {\cal O}(x_2) {\cal O}(x_3)  {\cal O}(x_4) \rangle \cr 
 &= { 4 \over N} \left( {1 \over |x_{12} x_{23} x_{34} x_{41}|^{2\Delta_\Phi}}  + {1 \over |x_{12} x_{24} x_{43} x_{31} |^{2\Delta_\Phi}} \right. \cr
  & \quad + \left. {1 \over |x_{13} x_{32} x_{24} x_{41}|^{2\Delta_\Phi}}\right),
 }
 where $4/N = 16 C^4 N$ with the factor of $16$ coming from the possible Wick contractions. Holographically, we found the four-point function to be given by \eno{FourPtFreeFinal}. Since $g_3$ has already been fixed, a non-trivial consistency check is to verify that \eno{g3Match} is consistent with
  \eqnQq{ConsistencyCond}{
  {4 \over N}= {g_3^2 \over 2} \widetilde{c}_{\Delta}^3 f_{\Delta\Delta\Delta}^2\,,
  }
and we find that it is. 
 
 In $\Rn$, the coupling constant for a spin-$\ell_1$--spin-$\ell_2$--spin-$\ell_3$ cubic vertex in the minimal bosonic higher spin theory conjecturally dual to the free $O(N)$ model in $n$ dimensions is (c.f.~(2.14) of~\cite{Bekaert:2015tva} for the scalar--scalar--spin-$\ell$ coupling, or more generally (1.12) of~\cite{Sleight:2016dba})
 \eqnRn{g3spinRn}{
 g_3^{(\ell_1,\ell_2,\ell_3)} &= \frac{\pi ^{\frac{n-3}{4}} 2^{\frac{1}{2} (3 n+\ell_1+\ell_2+\ell_3-1)}}{\sqrt{N}\: \Gamma_{\rm Euler}(n+\ell_1+\ell_2+\ell_3-3)} \cr
 & \quad \times \prod_{i=1}^3 \sqrt{\Gamma_{\rm Euler}(\ell_i + {n-1 \over 2}) \over \Gamma_{\rm Euler}(\ell_i+1)}\,,
 }
  which at $\ell_i=0$ for all $i$ reduces to the scalar--scalar--scalar coupling constant
  \eqnRn{g3scalarRn}{
  g_3^{(0,0,0)} = {1 \over \sqrt{N}} {2 \sqrt{2} \over \widetilde{c}_\Delta^{3/2} f_{\Delta\Delta\Delta}}\,,
  }
  where $\widetilde{c}_\Delta$ is given by \eno{ctDeltaRn}, $f_{\Delta\Delta\Delta}$ by \eno{fijk}, and $\Delta = n-2$. Equation \eno{g3scalarRn} is to be compared with the $p$-adic result in \eno{g3Match}. The authors of~\cite{Bekaert:2015tva} determine the full quartic scalar coupling in the bulk dual to the free $O(N)$ model by choosing an ansatz for the contact interaction schematically of the form   
  \eqnRn{quarticAnsatz}{
  {\cal V} &= \sum_{m,\ell=0}^\infty \lambda_{m,\ell} (\phi(x) \nabla_{\mu_1}\ldots \nabla_{\mu_\ell} \phi(x) + \cdots) \cr 
  & \qquad\times \square^m (\phi(x) \nabla^{\mu_1}\ldots \nabla^{\mu_\ell} \phi(x) + \cdots)\,,
  }
  which together with contributions from exchange diagrams~\cite{Bekaert:2014cea}, must reproduce the full connected four-point function of the free-theory. This leads to a generating function for the constants $\lambda_{m,\ell}$~\cite{Bekaert:2015tva}.

In the minimal construction presented in this paper, we have avoided a discussion of higher spin operators in the boundary theory since local currents in $p$-adic field theories are still not properly understood. Analogously, an understanding of gauge fields on the Bruhat--Tits tree remains elusive so far. With this caveat in mind, we may summarise the findings of this section in the form of a bulk quartic coupling on the Bruhat--Tits tree
 \eqnQq{quarticQq}{
 {\cal V} = \sum_{m=0}^1 \lambda_{m} \phi_a^2 \square^m \phi_a^2\,,
 }
 which, together with exchange diagrams coming from the cubic coupling reproduces the $O(1/N)$ four-point function of the free $p$-adic $O(N)$ model. Here $\square$ is the Laplacian on the tree, defined by
 \eqnQq{TreeLaplacian}{
 \square \phi_a  \equiv \sum_{b \sim a} (\phi_a - \phi_b)\,,
 }
where the sum $\sum_{b \sim a}$ is over the nearest neighbors $b$ of $a$, and the coefficients $\lambda_m$ are related to $g_4, \widetilde{g}_4$ given in \eno{CouplingsFree} via
 \eqnQq{lambdaVals}{
 \lambda_0 = {g_4 \over 4!} + {\widetilde{g}_4 \over 8} (p^n+1) \qquad \lambda_1 = - {\widetilde{g}_4 \over 8}\,.
 }
 In this section we were guided by Occam's razor to find the simplest action which produces the desired correlators of a free field theory, up to the four-point function. 
A reasonable expectation is that if we allow more interaction terms (for example, next-to-nearest neighbor quartic interactions), the bulk theory will no longer be entirely constrained by its correlators up to four-point functions.
Indeed, it is possible that the introduction of gauge degrees of freedom, or considerations of higher-point correlators, will suggest the existence of additional interaction vertices. It would be interesting to find symmetry principles which fully dictate the form of the bulk dual of the free $p$-adic $O(N)$ model.

\section{Discussion}
\label{DISCUSSION}

Despite appearances, $p$-adic AdS/CFT is not disconnected from the usual AdS/CFT correspondence; this becomes strikingly transparent when all quantities are expressed in terms of the right functions, namely the local zeta functions defined in \eno{zetaQq}-\eno{zetaRn}. In continuation of previous work~\cite{Gubser:2016guj}, we have shown in this paper explicit evidence in line with this point of view, via the holographic computation of the structure constants as well as the complete four-point function of scalar operators.  A source of considerable simplifications in $p$-adic field theories 
is the absence of derivatives in the $p$-adic OPE and hence in the conformal block decomposition of the $p$-adic four-point function. We believe that further insight into how $p$-adic and Archimedean field theories are to be compared may be gained by studying the integral representation of the $p$-adic OPE, which does not rely on a derivative expansion even in $\Rn$. Additionally, the simple structure of the OPE, and the remarkable similarity with the geodesic bulk diagram story in the Archimedean place, leads naturally to the expectation that a $p$-adic analog of kinematic space technology \cite{Czech:2016xec} exists. It would be interesting to explore this connection further.

Since the absence of derivatives leads to remarkable simplifications in calculations, we anticipate the computation of higher-point correlators as well as the evaluation of loop corrections in AdS to be considerably easier than in $\Rn$. It will be interesting to compute loop corrections and compare with recent results in $\Rn$~\cite{Aharony:2016dwx}. It will also be interesting to compute diagrams which have remained out of reach so far on the Archimedean side, since the $p$-adic results may potentially shed some light into computations in $\Rn$. To this end we have presented in appendix \ref{PROPIDENTITIES} some propagator identities  which we expect to be of use in brute-force computations of certain diagrams. 

Crossing symmetry and Mellin space methods were found to be especially useful in computing loop diagrams in $\Rn$~\cite{Aharony:2016dwx}. In $p$-adic CFTs, crossing symmetry is not as constraining as in $\Rn$; it merely restricts the OPE coefficients to obey the associativity property of a commutative algebra. Once the OPE coefficients have been chosen to obey the associativity property \eno{OPEAssoc}, there are no further constraints to impose on the scaling dimensions of operators.  On the other hand, application of Mellin space methods in the context of $p$-adic AdS/CFT may lead to further unexpected simplifications. Recent progress along the lines of~\cite{Gopakumar:2016cpb} would also be interesting to realize in the $p$-adic setting.
 
So far we have restricted ourselves to correlators of external operators without spin. It would be very interesting to include spin degrees of freedom both in the bulk and on the boundary. This will likely involve the use of more general multiplicative characters of the multiplicative group $\Qq^\times$, along the lines mentioned in~\cite{Heydeman:2016ldy,Gubser:2017vgc}. Comparisons with recent work on geodesic bulk diagrams for operators with spin~\cite{Hijano:2015zsa, Nishida:2016vds, Castro:2017hpx, Dyer:2017zef, Chen:2017yia} as well as other alternative approaches to conformal block decomposition, such as the one in \cite{Sleight:2017fpc}, would also be very interesting.
 
A difficulty in the study of $p$-adic AdS/CFT has been the absence of a clean dual pair, where on the bulk side we have a classical theory on the Bruhat--Tits tree and on the boundary side we have a large $N$ field theory which can be formulated independently of any holographic considerations.  Our calculations in section~\ref{FREECFT} bring us a step closer to exhibiting such a pair, as we summarize in the next two paragraphs.

On the field theory side, we have the free $O(N)$ model, which admits a lagrangian treatment and has deformations that lead to a Wilson-Fisher fixed point: See for example \cite{Missarov:2006iu,Gubser:2017vgc}.  On the bulk side, we have the theory \eno{Snearest} with couplings chosen as in \eno{CouplingsFree} and \eno{g3Match}.  Though this theory seems contrived, it has the virtue of matching the two-, three-, and four-point functions of the operator ${\cal O} = \vec\Phi^2$.  We should ask, what part of this matching was forced, or guaranteed, and what part is non-trivial?  The functional form of the two-point and three-point functions are fixed by conformal invariance, so that is an example of a guaranteed match once we choose the mass $m^2_\Delta$ correctly in the bulk action \eno{Snearest}.  The dependence of the four-point function on $u$ is {\it not} fixed by conformal invariance, but by including on-site cubic, on-site quartic, and nearest neighbor quartic interactions in \eno{Snearest}, we are giving ourselves just enough parameters to force an agreement in the functional form of the four-point function between the field theory and the bulk theory.  This agreement of the functional form of the four-point function is guaranteed once we impose the relations \eno{CouplingsFree}.

With functional forms matching perfectly between the field theory and the bulk, there remains the question of whether normalizations match.  At the level of our analysis, the normalization of the two-point function is another forced match, based essentially on choosing the normalization of the operator ${\cal O}$.  Even the normalization of the three-point function is a forced match, because we have one last free parameter in the bulk theory to adjust, namely the cubic coupling $g_3$.  The choice made in \eno{g3Match} guarantees a match in the normalization of the three-point function.  But there is one more calculation to do, namely the normalization of the four-point function!  The condition \eno{ConsistencyCond} providing for a precise match in the four-point normalization is non-trivial because all quantities involved in it have been fixed by previous considerations as just described.  Thus, finding that \eno{ConsistencyCond} holds is the first non-trivial match we have found between explicit field theory calculations and bulk calculations in $p$-adic AdS/CFT.

Of course, we hope for much more.  In particular, because the setup is so similar to the correspondence \cite{Klebanov:2002ja} between the Archimedean $O(N)$ model and Vasiliev theory in $AdS_4$ \cite{Vasiliev:1990en,Vasiliev:1992av}, we naturally hope to find some way to reformulate Vasiliev theory on a discrete geometry such as the Bruhat--Tits tree.  And we expect to see that the interacting Wilson-Fisher fixed point can be treated holographically just by changing boundary conditions on the bulk field $\phi$, as in the Archimedean case.  We hope to report on these and other related issues in future work.

\subsection*{Acknowledgments}

We thank B.~Czech, E.~Perlmutter, and S.~Sondhi for useful discussions. This work was supported in part by the Department of Energy under Grant No.~DE-FG02-91ER40671. The work of S.S.G.\ was performed in part at the Aspen Center for Physics, which is supported by National Science Foundation grant PHY-1066293.  The work of S.P.\ was also supported in part by the Bershadsky Family Fellowship Fund in Mathematics or Physics.

\appendix
\section{Some more propagator identities}
\label{PROPIDENTITIES}
In this section we list (without proof) some additional propagator identities on the Bruhat--Tits tree which could prove useful in evaluating higher-point correlators and higher loop bulk integrals. Moreover, it is likely they will be useful in obtaining by analogy with $\Qq$ the corresponding, as yet unknown, propagator identities in $\Rn$, which could in turn potentially simplify the evaluation of loop diagrams and higher-point functions in $\Rn$.

\subsection{Identities involving two propagators}
Two identities, similar in spirit to \eno{IntGGnormQq}, are:
\eqnQq{IntKK}{
\sum_{c \in T_{p^n}} &\hat{K}_{\Delta_1}(x_1,c) \hat{K}_{\Delta_2}(x_2,c) \cr 
&= {\hat{K}_{\Delta_1}(x_1,o) \hat{K}_{\Delta_2}(x_2,o)  \over m_{\Delta_1+\Delta_2}^2 \zeta(\Delta_1+\Delta_2)}\,,
}
where $o$ is any fixed bulk point lying on the geodesic between $x_1$ and $x_2$, and
\eqnQq{IntGK}{
\sum_{c \in T_{p^n}} {G}_{\Delta_1}(a_1,c) \hat{K}_{\Delta_2}(x_2,c) =   {\hat{K}_{\Delta_2}(x_2,a_1) \over m_{\Delta_1}^2 - m_{\Delta_2}^2 }\,.
}
\subsection{Identities involving three propagators}

We begin by recalling the three-point amplitude
\eqnQq{ThreePtKKK}{
& \sum_{c \in T_{p^n}} \hat{K}_{\Delta_1}(x_1,c)  \hat{K}_{\Delta_2}(x_2,c) \hat{K}_{\Delta_3}(x_3,c) \cr
& = \hat{K}_{\Delta_1}(x_1,o) \hat{K}_{\Delta_2}(x_2,o) \hat{K}_{\Delta_3}(x_3,o)\, f_{123}\,,
}
where $o$ is the unique point of intersection of the geodesics connecting the (boundary) points $x_1, x_2, x_3$, and $f_{123}$ is a constant given in \eno{StrConstQq}.

A few identities which may be useful in evaluating loop diagrams involve, similar to \eno{ThreePtKKK}, a reduction of the integration over a bulk point of a product of three propagators, to a combination of unintegrated propagators. We list here the identities:
\eqnQq{IntGKK}{
 & \sum_{c \in T_{p^n}}  {G}_{\Delta_1}(a_1,c) \hat{K}_{\Delta_2}(x_2,c) \hat{K}_{\Delta_3}(x_3,c) \cr
 &= {G}_{\Delta_1}(a_1,o) \hat{K}_{\Delta_2}(x_2,o) \hat{K}_{\Delta_3}(x_3,o)\, f_{123} \cr 
& \quad - {\hat{K}_{\Delta_2}(x_2,a_1) \hat{K}_{\Delta_3}(x_3,a_1) \over m_{\Delta_2+\Delta_3}^2-m_{\Delta_1}^2}\,,
}
where $o$ is now the bulk point of intersection of geodesics connecting $a_1, x_2, x_3$. In addition,
\eqnQq{IntGGK}{
& \sum_{c \in T_{p^n}}  {G}_{\Delta_1}(a_1,c) {G}_{\Delta_2}(a_2,c) \hat{K}_{\Delta_3}(x_3,c) \cr
& = {G}_{\Delta_1}(a_1,o) {G}_{\Delta_2}(a_2,o) \hat{K}_{\Delta_3}(x_3,o)\, f_{123} \cr 
& \quad - {{G}_{\Delta_2}(a_2,a_1) \hat{K}_{\Delta_3}(x_3,a_1) \over m_{\Delta_2+\Delta_3}^2-m_{\Delta_1}^2} \cr 
& \quad - {{G}_{\Delta_1}(a_1,a_2) \hat{K}_{\Delta_3}(x_3,a_2) \over m_{\Delta_1+\Delta_3}^2-m_{\Delta_2}^2} \,,
}
and
\eqnQq{IntGGG}{
&\sum_{c \in T_{p^n}} {G}_{\Delta_1}(a_1,c) {G}_{\Delta_2}(a_2,c) {G}_{\Delta_3}(a_3,c) \cr
& = {G}_{\Delta_1}(a_1,o) {G}_{\Delta_2}(a_2,o) {G}_{\Delta_3}(a_3,o)\, f_{123} \cr 
& \quad - {{G}_{\Delta_2}(a_2,a_1) {G}_{\Delta_3}(a_3,a_1) \over m_{\Delta_2+\Delta_3}^2-m_{\Delta_1}^2} \cr 
& \quad - {{G}_{\Delta_1}(a_1,a_2) {G}_{\Delta_3}(a_3,a_2) \over m_{\Delta_1+\Delta_3}^2-m_{\Delta_2}^2} \cr 
& \quad - {{G}_{\Delta_1}(a_1,a_3) {G}_{\Delta_2}(a_2,a_3) \over m_{\Delta_1+\Delta_2}^2-m_{\Delta_3}^2}\,.
}
 As a check, when $a_1 = a_2$ (in which case the point of intersection $o=a_1=a_2$), we recover from \eno{IntGGG} the two propagator identity \eno{IntGGnormQq},
\eqnQq{IntThreeGReduce}{
\sum_{c \in T_{p^n}}& {G}_{\Delta_1+\Delta_2}(a_1,c)  {G}_{\Delta_3}(a_3,c) \cr 
& = {{G}_{\Delta_1+\Delta_2}(a_1,a_3)  - {G}_{\Delta_3}(a_1,a_3) \over m_{\Delta_3}^2 - m_{\Delta_1+\Delta_2}^2}\,.
}

\section{Crossing symmetry of the four-point function}
\label{CROSSING}
In this appendix, we demonstrate how crossing symmetry of the $p$-adic four-point function works with an explicit example (see~\cite{Melzer1989} for a more general argument). Before we begin, it is useful to emphasize and clarify the notation used in this paper, since it can lead to some confusion. Throughout the paper, we have assumed that the boundary insertion points $x_i$ are in a configuration as depicted in figure \ref{figSchannel}, i.e.~with $u< v=1$, and we have reserved the term `$s$-channel' to refer to that. (This admittedly confusing terminology has no connection with the textbook terminology for exchange of particles in an intermediate channel. Instead we refer to exchanges as, for example, the exchange in the $(12)(34)$ channel.) Note that this is not a simplification but in fact very general on the Bruhat--Tits tree. Given any four boundary points, up to relabelling the $x_i$, they always arrange themselves in an `$s$-channel' configuration, with one exception. The exceptional case corresponds to $u=v=1$, or equivalently  when the bulk points $c_1$ and $c_2$ in figure \ref{figSchannel} coincide. All our formulae derived in this paper are applicable when $u=v=1$, which can be thought of as a degeneration of the $s$-, $t$-, and $u$-channel boundary configurations, where for example, the `$t$-channel' configuration is  obtained after switching $x_2$ and $x_3$ in figure \ref{figSchannel}.\footnote{Curiously, this degeneration is impossible when $p^n = 2$, since the Bruhat--Tits tree $T_2$ has coordination number $3$, while we need a coordination number of at least $4$ to realize $u=v=1$. This is one of the many reasons, which are all related to the fact that $2$ is an even prime, why $2$-adic conformal field theories may be quite exotic.} It is worth emphasizing that $u=v=1$ is the {\it only} case which admits a pairwise overlap between the $s$-, $t$- and $u$-channels~\cite{Melzer1989}.

Now let us consider an exchange diagram where the boundary configuration is in the `$s$-channel' (more accurately, we'd like to allow $u \leq v = 1$), and a scalar of dimension $\Delta$ is exchanged in the $(13)(24)$ channel. This is precisely the diagram defined in \eno{ExchangeDiagCrossed}, whose conformal block decomposition in the crossed channel is given in \eno{ExchTFinal}-\eno{OPEsqExchT34}. As expected it involves just the conformal blocks of double-trace operators. If we wanted the conformal block decomposition of \eno{ExchangeDiagCrossed} in the direct channel (for which the boundary points $x_i$ must be able to admit a `$t$-channel' configuration as well), we need only adapt the result in \eno{ExchSFinal}-\eno{OPEsqExch} by making the replacements $\Delta_2 \leftrightarrow \Delta_3$ and $x_2 \leftrightarrow x_3$. (This prescription is obvious upon comparing \eno{ExchDiagS} with \eno{ExchangeDiagCrossed}.) This leads to
 \eqnQq{ExchDiagTDirect}{
 {\cal D}_\Delta^T &=  P_1{(13)} {\cal G}_{\Delta_C}(u^\prime) + P_1{(24)} {\cal G}_{\Delta_D}(u^\prime) \cr
 & \quad + C_{13 \Delta} C_{24 \Delta} {\cal G}_\Delta(u^\prime)
 }
 where the OPE coefficients-squared are obtained by making the replacements in \eno{ExchSFinal}-\eno{OPEsqExch} as described above, $\Delta_C, \Delta_D$ are defined in \eno{DeltaCD} and 
 \eqnQq{uPrimeDef}{
 u^\prime  = \left| { x_{13} x_{24} \over x_{12} x_{34}} \right| = {1 \over u}\,.
 }
 As expected in the direct channel, in addition to the double-trace operators, the single-trace operator exchanged in the intermediate channel appears in the conformal block decomposition.
 Now crossing symmetry requires that the expressions in the direct and crossed channels agree. The crucial point is that  we required the boundary points $x_i$ to admit both an `$s$-channel' as well a `$t$-channel' configuration. As remarked earlier, this forces $u = 1/u^\prime = 1$ (and $v$ remains fixed at $v=1$). Plugging this in the expressions and comparing \eno{ExchTFinal}-\eno{OPEsqExchT34} with \eno{ExchDiagTDirect}, we find they agree exactly as required by crossing symmetry.

\section{Cubic derivative interactions}
\label{CUBICDERIV}
In this appendix we briefly discuss  bulk cubic couplings with derivatives. We have shown in section \ref{PADICOPE} that a $\phi_1\phi_2\phi_3$ bulk coupling results in a three-point amplitude for the dual operator given by
 \eqnQq{ThreePtAmp}{
 \mathcal{A} &\equiv {1\over V(x_i)}\sum_{a \in T_{p^n}} \prod_i^3 \hat{K}_{\Delta_i}(x_i,a) \cr
 &= f_{123}\,,
 }
 where $f_{ijk}$ is given in \eno{fijk} and
\eqn{VDef}{
& V(x_i) 
 = {1 \over |x_{12}|^{\Delta_1+\Delta_2-\Delta_3} |x_{23}|^{\Delta_2+\Delta_3-\Delta_1} |x_{13}|^{\Delta_3+\Delta_1-\Delta_2}}.
}
The corresponding amplitude in $\Rn$ is~\cite{Freedman:1998tz}
 \eqnRn{ThreePtAmpRn}{
 \mathcal{A} &\equiv {1\over V(x_i)} \int {d^{n+1} y\over y_0^{n+1}} \prod_i^3 \hat{K}_{\Delta_i}(y_0,\vec{y}-\vec{x}_i) \cr
 &= {1 \over 2}f_{123}\,.
 }
 
We now describe what a derivative cubic coupling, schematically of the form  $\phi (\nabla \phi)^2$, looks like on the Bruhat--Tits tree.  Mimicking $\Rn$, where such a coupling arises from a $\phi_1 g^{\mu\nu} \partial_\mu \phi_2 \partial_\nu \phi_3$ interaction vertex, we posit an obvious candidate vertex on the Bruhat--Tits tree,
 \eqnQq{CubicDerivative}{
 S \supset
 \sum_{\langle ab\rangle} \phi_{1a} \left(\phi_{2a}-\phi_{2b}\right) \left(\phi_{3a}-\phi_{3b}\right),
 }
where $\phi_i$ has scaling dimension $\Delta_i$ and $a,b$ label vertices on the tree. The symbol $\sum_{\langle ab \rangle}$ stands for summing over all pairs of nearest neighbors. It is apparent from the structure of \eno{CubicDerivative} that it involves nearest-neighbor interactions of the form $\phi_a \phi_a \phi_b$, where $a$ and $b$ are adjacent vertices on the tree. Thus introducing derivative bulk couplings is tantamount to introducing nearest-neighbor interactions on the tree. (More generally, for higher derivative couplings, one should introduce (next)$^k$-to-nearest neighbor interactions on the tree.) Computing the amplitude arising from \eno{CubicDerivative}, we obtain
 \eqnQq{ThreePtDeriv}{
 \mathcal{A}_{\partial} &\equiv {1 \over V} \sum_{\langle ab \rangle}  \hat{K}_{\Delta_1}(x_1,a) ( \hat{K}_{\Delta_2}(x_2,a) - \hat{K}_{\Delta_2}(x_2,b)) \cr
 &\qquad \times ( \hat{K}_{\Delta_3}(x_3,a) - \hat{K}_{\Delta_3}(x_3,b)) \cr
 &= {1 \over 2} \left(m_{\Delta_1}^2 - m_{\Delta_2}^2  - m_{\Delta_3}^2 \right) \mathcal{A} \,,
 }
 where $\mathcal{A}$ is given by \eno{ThreePtAmp}. The final expression in \eno{ThreePtDeriv} follows from a straightforward computation on the Bruhat--Tits tree. 
 The corresponding amplitude in $\Rn$ is~\cite{Freedman:1998tz}
 \eqnRn{ThreePtDerivRn}{
 \mathcal{A}_{\partial} &\equiv {1 \over V}\int {d^{n+1} y \over y_0^{n+1}}\, \hat{K}_{\Delta_1}(y_0, \vec{y}-\vec{x}_1)   \cr
 &\quad \times \partial_\mu \hat{K}_{\Delta_2}(y_0, \vec{y}-\vec{x}_2) y_0^2 \partial^\mu \hat{K}_{\Delta_3}(y_0, \vec{y}-\vec{x}_3)  \cr 
 & = {1 \over 2}\left(m_{\Delta_1}^2 - m_{\Delta_2}^2  - m_{\Delta_3}^2 \right) \mathcal{A} \,,
 }
where $\mathcal{A}$ is given by \eno{ThreePtAmpRn}.  The identical form of the amplitudes in \eno{ThreePtDeriv} and \eno{ThreePtDerivRn} supports the claim that \eno{CubicDerivative} is indeed a derivative coupling on the tree.

\section{Direct computation of geodesic bulk diagrams}
\label{MOREDETAILS}
Using variants of \eno{KKaQq} and \eno{IntGGQq} repeatedly, we can convert the four-point amplitudes (such as those in \eno{Dfn}, \eno{ExchDiagS} and \eno{ExchangeDiagCrossed}) into a sum of several terms of the form given in  \eno{GeoBulkDiag} and \eno{GeodesicbulkDiagCrossed}, repeated below for convenience:
  \eqn{Wsums}{
  {\cal W}_\Delta^S &= \sum_{b_1 \in \gamma_{12} \atop b_2 \in \gamma_{34}} 
   (x_1 b_1)^{\Delta_1} (x_2 b_1)^{\Delta_2} (b_1 b_2)^\Delta
    (x_3 b_2)^{\Delta_3} (x_4 b_2)^{\Delta_4}  \cr
    {\cal W}_\Delta^T &= \sum_{b_1 \in \gamma_{13} \atop b_2 \in \gamma_{24}} 
   (x_1 b_1)^{\Delta_1} (x_3 b_1)^{\Delta_3} (b_1 b_2)^\Delta
    (x_2 b_2)^{\Delta_2} (x_4 b_2)^{\Delta_4}  \cr
     {\cal W}_\Delta^U &= \sum_{b_1 \in \gamma_{14} \atop b_2 \in \gamma_{23}} 
   (x_1 b_1)^{\Delta_1} (x_4 b_1)^{\Delta_4} (b_1 b_2)^\Delta
    (x_2 b_2)^{\Delta_2} (x_3 b_2)^{\Delta_3} \,.
 }
(One such computation was detailed in section \ref{EXCHANGE}.) Here we write $(ab)^\Delta$ instead of $\hat{G}_\Delta(b,a)$ or $\hat{K}_\Delta(b,a)$ for brevity. The symbol $\gamma_{ij}$ denotes a geodesic joining boundary points $x_i$ and $x_j$. From here on we will denote geodesic paths by $(x_i:x_j)$. If paths from boundary points $x_1$, $x_2$, and $x_3$ meet at a bulk point $c$, then we have
 \eqnQq{Kcx}{
  (x_1 c)^\Delta = \left| {x_{23} \over x_{12} x_{13}} \right|^\Delta \,.
 }
We restrict attention to configurations of the $x_i$ such that $u \leq 1$ and $v = 1$, where $u$ and $v$ are defined in \eno{uvDef}. We can do this without loss of generality because upto relabelling of $x_i$, we can always arrange $u \leq 1$ and $v=1$.
In these sorts of configurations, paths from $x_1$ and $x_2$ converge at a bulk point $c_1$, which connects to a bulk point $c_2$ where paths from $x_3$ and $x_4$ converge (see, for example, figure \ref{figSchannel}).  An important identity is
 \eqnQq{dcu}{
  u = \left|{ x_{12} x_{34} \over x_{13}x_{24}} \right| =  (c_1 c_2) = p^{-d(c_1,c_2)} \,.
 }
If $u=1$ then $c_1$ and $c_2$ coincide.

 In the main text, we noted without proof that the sums in \eno{Wsums}, which are referred to as geodesic bulk diagrams, are related to $p$-adic conformal blocks via the identities in \eno{GeodesicConfWaveQq} and \eno{GeodesicConfWaveT}. The goal of this appendix is to prove these identities by direct computation.

  We can conveniently factor out most of the $x_i$ dependence from any of the geodesic bulk diagrams we consider by dividing out by the trivial conformal partial wave $W_0(x_i)$, given in \eno{WZeroExpressQq}. A more convenient form for $W_0$ follows from \eno{WDeltaDef}, 
  \eqnQq{W0again}{
  W_0(x_i) = (x_1c_1)^{\Delta_1}(x_2c_1)^{\Delta_2}(x_3c_2)^{\Delta_3}(x_4c_2)^{\Delta_4}\,.
  }
   Then for any of the geodesic bulk diagrams ${\cal W}$ we consider, ${\cal W} / W_0(x_i)$ is a function of the $x_i$ only through dependence on the cross-ratio $u$. An easy case is 
 \begin{widetext}
  \eqnQq{GsSum}{
  {{\cal W}_\Delta^S \over W_0} &= \left[ \sum_{b_1 \in (x_1:c_1]} 
    {(x_1 b_1)^{\Delta_1} (x_2 b_1)^{\Delta_2} (b_1 c_1)^\Delta \over 
     (x_1 c_1)^{\Delta_1} (x_2 c_1)^{\Delta_2}} + 
    \sum_{b_1 \in (c_1:x_2)} 
    {(x_1 b_1)^{\Delta_1} (x_2 b_1)^{\Delta_2} (b_1 c_1)^\Delta \over 
     (x_1 c_1)^{\Delta_1} (x_2 c_1)^{\Delta_2}} \right] u^\Delta  \cr
   &\qquad{} \times \left[ \sum_{b_2 \in (x_3:c_2]} 
    {(x_3 b_2)^{\Delta_3} (x_4 b_2)^{\Delta_4} (b_2 c_2)^\Delta \over 
     (x_3 c_2)^{\Delta_3} (x_4 c_2)^{\Delta_4}} + 
    \sum_{b_2 \in (c_2:x_4)} 
    {(x_3 b_2)^{\Delta_3} (x_4 b_2)^{\Delta_4} (b_2 c_2)^\Delta \over 
     (x_3 c_2)^{\Delta_3} (x_4 c_2)^{\Delta_4}} \right].
 }
 	\begin{center}
 	\begin{figure}
 	\begin{overpic}[width=0.8\textwidth]{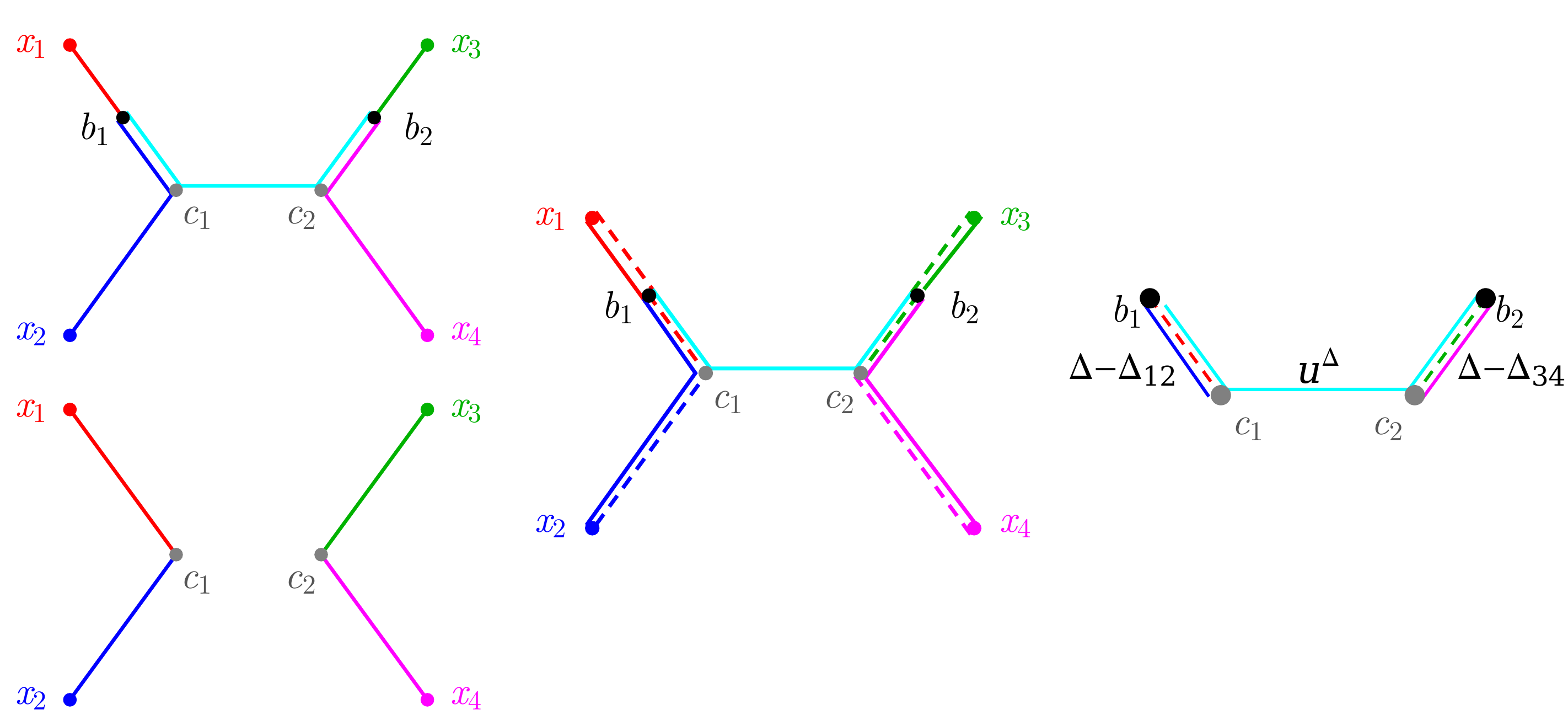}
	 \put (33,21) {\LARGE =}
	 \put (63,21) {\LARGE =}
	 \put (0,22) {\line(1,0){31}}
	\end{overpic}
	 \caption{(Color online.) A graphical method of representing terms in \eno{GsSum}-\eno{FsAgain}.  Here and in figure~\ref{fig:WtDiagrams}, a solid line means that the amplitude should include a factor of the propagator between the endpoints of that line, whereas a dashed line means that we are dividing by that propagator.  When a combination of solid and dashed lines is labeled with a power $\delta$, like $\Delta-\Delta_{12}$, it means that each step along this combinations of lines is weighted by a factor of $p^{-\delta}$.  To improve readability we use $u^\delta$ instead of $\delta$ to label combined lines between $c_1$ and $c_2$.\label{fig:WsDiagram}}
	 \end{figure}
	\end{center}

 	\begin{center}
 	\begin{figure}
 	\begin{overpic}[width=0.8\textwidth]{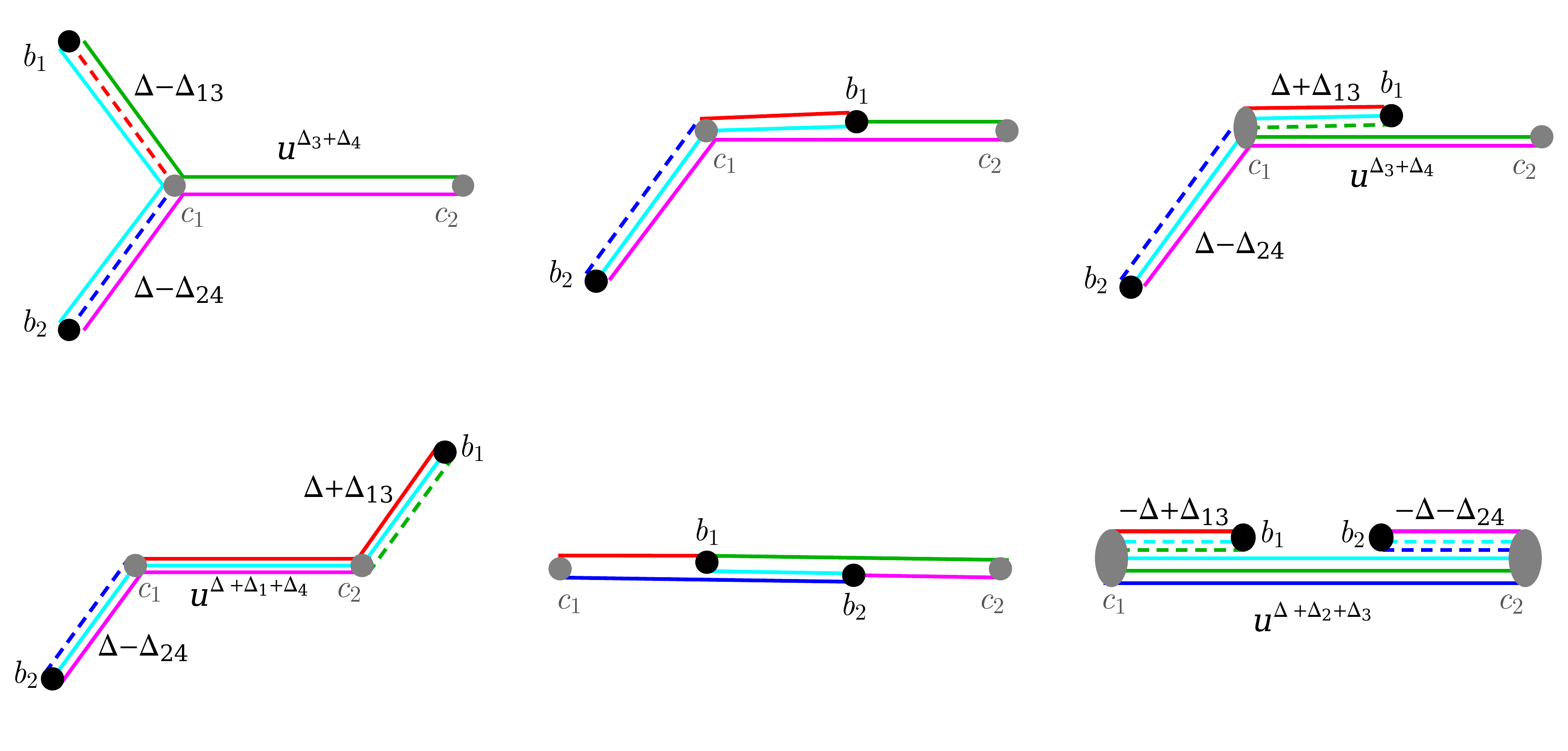}
	 \put (66,34) {\LARGE =}
	 \put (66,11) {\LARGE =}
	 \put (-3,40) {\LARGE \raisebox{.5pt}{\textcircled{\raisebox{-.9pt} {$1$}}}}
	 \put (34,40) {\LARGE \raisebox{.5pt}{\textcircled{\raisebox{-.9pt} {$2$}}}}
	 \put (-3,17) {\LARGE \raisebox{.5pt}{\textcircled{\raisebox{-.9pt} {$3$}}}}
	 \put (34,17) {\LARGE \raisebox{.5pt}{\textcircled{\raisebox{-.9pt} {$4$}}}}
	\end{overpic}
	 \caption{(Color online.) Subway diagrams leading to summands in \eno{bbPossibilities}. \label{fig:WtDiagrams}}
	 \end{figure}
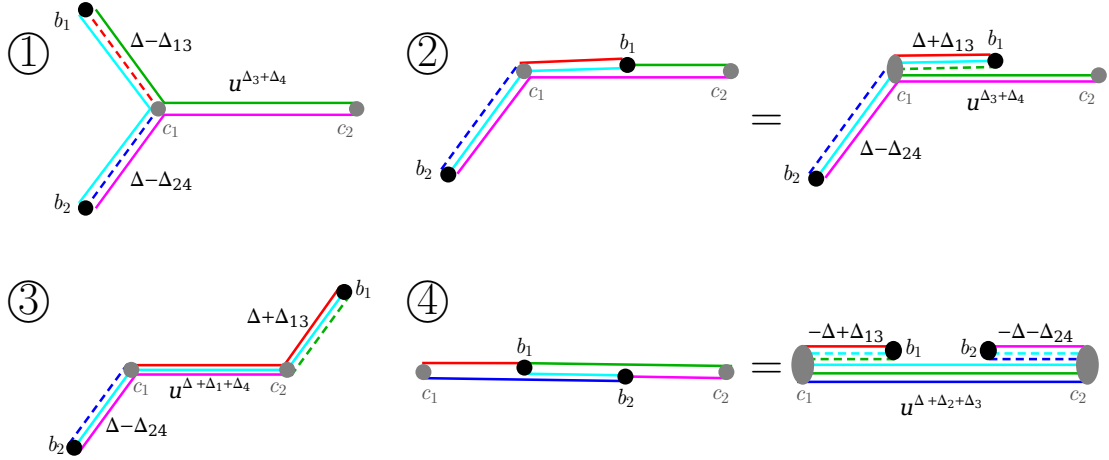
	\end{center}
 \end{widetext}
What makes this case relatively easy is that $b_1$ and $b_2$ cannot belong to $(c_1:c_2)$, because the paths $(x_1:x_2)$ and $(x_3:x_4)$ do not have any points in common with $(c_1:c_2)$.  Using identities like $(x_1 c_1) = (x_1 b_1) (b_1 c_1)$ and $(x_2 b_1) = (x_1 c_1) (b_1 c_1)$, we may simplify \eno{GsSum} to
 \eqnQq{FsAgain}{
  {{\cal W}_\Delta^S \over W_0} &= \!u^\Delta \!\!\left[ \sum_{b_1 \in (x_1:c_1]}\!\!\!\! (b_1 c_1)^{\Delta-\Delta_{12}} + 
    \!\!\!\!\sum_{b_1 \in (c_1:x_2)}\!\!\!\! (b_1 c_1)^{\Delta+\Delta_{12}} \right]   \cr
   & \!\!\times
    \left[ \sum_{b_2 \in (x_3:c_2]}\!\!\!\! (b_2 c_2)^{\Delta-\Delta_{34}} +
    \!\!\!\!\sum_{b_2 \in (c_2:x_4)}\!\!\!\! (b_2 c_2)^{\Delta+\Delta_{34}} \right]  \cr
   &= \beta^{(\Delta+\Delta_{12}, \Delta-\Delta_{12})} \beta^{(\Delta+\Delta_{34},\Delta-\Delta_{34})} \,,
 }
 where $\beta^{(s,t)}$ is defined in \eno{betaDef}.
 This proves \eno{GeodesicConfWaveQq}.
A graphical method of obtaining \eno{FsAgain} is shown in figure~\ref{fig:WsDiagram}.  The diagram furthest to the right in this figure corresponds to choosing the first term in both sets of square brackets in \eno{FsAgain}.  It is worth noting that the sums in \eno{FsAgain} converge iff the four quantities $\Delta \pm \Delta_{12}$ and $\Delta \pm \Delta_{34}$ are all positive.  If any one of them goes to $0$, then we have a logarithmic divergence.

${\cal W}_\Delta^T$ is more complicated than ${\cal W}_\Delta^S$ because the paths $(x_1:x_3)$ and $(x_2:x_4)$ are longer than $(x_1:x_2)$ and $(x_3:x_4)$, and there are more distinct ways to position $b_1$ and $b_2$ along these paths.  The full list of possibilities can be enumerated as follows:
 \eqn{bbPossibilities}{\seqalign{\span\TC & \;\span\TC & \;\;\span\TC & \;\;\span\TC}{
  \# & b_1 \in\qquad & b_2 \in\qquad & \hbox{summand}  \cr\hline
  1\colon & (x_1:c_1) & (x_2:c_1) & 
    u^{\Delta_B} (b_1 c_1)^{\Delta-\Delta_{13}} (b_2 c_1)^{\Delta-\Delta_{24}}  \cr 
  \tilde{1}\colon & (x_3:c_2) & (x_4:c_2) &
    u^{\Delta_A} (b_1 c_2)^{\Delta-\Delta_{13}} (b_2 c_2)^{\Delta+\Delta_{24}}  \cr
  2\colon & [c_1:c_2] & (x_2:c_1) & 
    u^{\Delta_B} (b_1 c_1)^{\Delta+\Delta_{13}} (b_2 c_1)^{\Delta-\Delta_{24}}  \cr
  \tilde{2}\colon & [c_1:c_2] & (x_4:c_2) &
    u^{\Delta_A} (b_1 c_2)^{\Delta-\Delta_{13}} (b_2 c_2)^{\Delta+\Delta_{24}}  \cr
  \underline{2}\colon & (x_1:c_1) & [c_1:c_2] & 
    u^{\Delta_B} (b_1 c_1)^{\Delta+\Delta_{24}} (b_2 c_1)^{\Delta-\Delta_{13}}  \cr
  \underline{\tilde{2}}\colon & (x_3:c_2) & [c_1:c_2] &
    u^{\Delta_A} (b_1 c_2)^{\Delta-\Delta_{24}} (b_2 c_2)^{\Delta+\Delta_{13}}  \cr 
  3\colon & (x_3:c_2) & (x_2:c_1) & 
    u^{\Delta+\Delta_E} (b_1 c_2)^{\Delta+\Delta_{13}} (b_2 c_2)^{\Delta-\Delta_{24}}  \cr
  \tilde{3}\colon & (x_2:c_1) & (x_4:c_2) &
    u^{\Delta+\Delta_F} (b_1 c_1)^{\Delta-\Delta_{13}} (b_2 c_1)^{\Delta+\Delta_{24}}  \cr
  4\colon & [c_1:c_2] & [b_1:c_2] & 
    u^{\Delta+\Delta_F} (b_1 c_1)^{-\Delta+\Delta_{13}} 
      (b_2 c_2)^{-\Delta-\Delta_{24}}  \cr
  \tilde{4}\colon & (c_1:c_2] & [c_1:b_1) & 
    u^{\Delta+\Delta_E} (b_1 c_2)^{-\Delta-\Delta_{13}} (b_2 c_1)^{-\Delta+\Delta_{24}}
 }}
 where $\Delta_A, \Delta_B, \Delta_E$ and $\Delta_F$ are defined in footnote \ref{fn:assoc}.
The summands can be written down by inspection of the relevant subway diagrams, a representative sampling of which is shown in figure~\ref{fig:WtDiagrams}.

Of the ten rows of \eno{bbPossibilities}, only the following five need to be computed explicitly:
 \eqn{Vone}{
  V_1 &\equiv u^{\Delta_B} \sum_{b_1 \in (x_1:c_1) \atop b_2 \in (x_2:c_1)}
     (b_1 c_1)^{\Delta-\Delta_{13}} (b_2 c_1)^{\Delta-\Delta_{24}}  \cr
   &= u^{\Delta_B} 
    \left[ \sum_{m_1=1}^\infty p^{-m_1(\Delta-\Delta_{13})} \right]
    \left[ \sum_{m_2=1}^\infty p^{-m_2(\Delta-\Delta_{24})} \right]
 }
 \eqn{Vtwo}{
  V_2 &\equiv u^{\Delta_B} \sum_{b_1 \in [c_1:c_2] \atop b_2 \in (x_2:c_1)}
     (b_1 c_1)^{\Delta+\Delta_{13}} (b_2 c_1)^{\Delta-\Delta_{24}}  \cr
   &= u^{\Delta_B}
    \left[ \sum_{m_1=0}^d p^{-m_1(\Delta+\Delta_{13})} \right]
    \left[ \sum_{m_2=1}^\infty p^{-m_2(\Delta-\Delta_{24})} \right]
 }
 \eqn{Vthree}{
  V_3 &\equiv u^{\Delta+\Delta_E} \sum_{b_1 \in (x_3:c_2) \atop b_2 \in (x_2:c_1)} 
     (b_1 c_2)^{\Delta+\Delta_{13}} (b_2 c_2)^{\Delta-\Delta_{24}}  \cr
   &= u^{\Delta+\Delta_E} \left[ \sum_{m_1=1}^\infty p^{-m_1(\Delta+\Delta_{13})} \right]
    \left[ \sum_{m_2=1}^\infty p^{-m_2(\Delta-\Delta_{24})} \right]
 }
 \eqn{Vfour}{
  V_4 &\equiv u^{\Delta+\Delta_F} \sum_{b_1 \in [c_1:c_2] \atop b_2 \in [b_1:c_2]}
    (b_1 c_1)^{-\Delta+\Delta_{13}} (b_2 c_2)^{-\Delta-\Delta_{24}}  \cr
   &= u^{\Delta+\Delta_F} \sum_{m_1=0}^d p^{-m_1(-\Delta+\Delta_{13})}
     \sum_{m_2=0}^{d-m_1} p^{-m_2(-\Delta-\Delta_{24})}
 }
 \eqn{Vfourt}{
  V_{\tilde{4}} &\equiv u^{\Delta+\Delta_E} \sum_{b_1 \in (c_1:c_2] \atop
      b_2 \in [c_1:b_1)} (b_1 c_2)^{-\Delta-\Delta_{13}} (b_2 c_1)^{-\Delta+\Delta_{24}}  \cr
    &= u^{\Delta+\Delta_E} \sum_{m_1=0}^{d-1} p^{-m_1(-\Delta-\Delta_{13})}
      \sum_{m_2=0}^{d-1-m_1} p^{-m_2(-\Delta+\Delta_{24})}\,,
 }
 where $d = d(c_1,c_2) = -\log_p u$.
Using obvious relations like
 \eqnQq[c]{ObviousRelations}{
  \sum_{m=0}^\infty p^{-ma} = \zeta(a) \qquad
  \sum_{m=1}^\infty p^{-ma} = -\zeta(-a)   \cr
  \sum_{m=0}^d p^{-ma} = \zeta(a) + u^a \zeta(-a) \cr
  \zeta(a) \zeta(b) + \zeta(a+b) \zeta(-b) - \zeta(a+b) \zeta(a) = 0 \,,
 }
we can simplify \eno{Vone}-\eno{Vfourt} to
 \eqnQq{FirstFive}{
  V_1 &= u^{\Delta_3+\Delta_4} \zeta(-\Delta+\Delta_{13}) \zeta(-\Delta+\Delta_{24})  \cr
  V_2 &= -u^{\Delta_3+\Delta_4} \zeta(\Delta+\Delta_{13}) \zeta(-\Delta+\Delta_{24}) \cr 
     &\quad - u^{\Delta+\Delta_1+\Delta_4} \zeta(-\Delta-\Delta_{13}) \zeta(-\Delta+\Delta_{24})  \cr
  V_3 &= u^{\Delta+\Delta_1+\Delta_4} \zeta(-\Delta-\Delta_{13}) \zeta(-\Delta+\Delta_{24})  \cr
  V_4 &= u^{\Delta_1+\Delta_2} \zeta(\Delta-\Delta_{13}) \zeta(-\Delta_{13}-\Delta_{24}) \cr
  & \quad + u^{\Delta_3+\Delta_4} \zeta(\Delta+\Delta_{24}) \zeta(\Delta_{13}+\Delta_{24})  \cr
   &\quad + u^{\Delta+\Delta_2+\Delta_3} \zeta(-\Delta+\Delta_{13}) \zeta(-\Delta-\Delta_{24}) \cr
  V_{\tilde{4}} &= 
    -u^{\Delta_1+\Delta_2} \zeta(-\Delta+\Delta_{24}) \zeta(-\Delta_{13}-\Delta_{24}) \cr
    & \quad - u^{\Delta_3+\Delta_4} \zeta(-\Delta-\Delta_{13}) \zeta(\Delta_{13}+\Delta_{24})  \cr
    &\quad + u^{\Delta+\Delta_1+\Delta_4} \zeta(-\Delta-\Delta_{13}) \zeta(-\Delta+\Delta_{24})\,.
 }
To obtain the remaining five amplitudes explicitly, we can either swap $1 \leftrightarrow 3$ and $2 \leftrightarrow 4$ (e.g.~to go from $2$ to $\tilde{2}$) or $1 \leftrightarrow 2$ and $3 \leftrightarrow 4$ (to go from $2$ to $\underline{2}$).  In summary,
 \eqnQq{FiveMore}{
  V_{\tilde{1}} &= u^{\Delta_1+\Delta_2} \zeta(-\Delta-\Delta_{13}) \zeta(-\Delta-\Delta_{24})  \cr
  V_{\tilde{2}} &= -u^{\Delta_1+\Delta_2} \zeta(\Delta-\Delta_{13}) \zeta(-\Delta-\Delta_{24}) \cr 
  &- u^{\Delta+\Delta_2+\Delta_3} \zeta(-\Delta+\Delta_{13}) \zeta(-\Delta-\Delta_{24})  \cr
  V_{\underline{2}} &= -u^{\Delta_3+\Delta_4} \zeta(-\Delta+\Delta_{13}) 
     \zeta(\Delta+\Delta_{24}) \cr 
     &\quad -u^{\Delta+\Delta_2+\Delta_3} \zeta(-\Delta+\Delta_{13}) \zeta(-\Delta-\Delta_{24})  \cr
  V_{\underline{\tilde{2}}} &= -u^{\Delta_1+\Delta_2} \zeta(-\Delta-\Delta_{13})
     \zeta(\Delta-\Delta_{24}) \cr
     &\quad -u^{\Delta+\Delta_1+\Delta_4} \zeta(-\Delta-\Delta_{13}) \zeta(-\Delta+\Delta_{24})  \cr
  V_{\tilde{3}} &= u^{\Delta+\Delta_2+\Delta_3} \zeta(-\Delta+\Delta_{13})
     \zeta(\Delta-\Delta_{24})\,.
 }
Our eventual goal is to add all ten $V_a$ together.  As a step in that direction, let's call $u^{\Delta_1+\Delta_2}$ and $u^{\Delta_3+\Delta_4}$ ``compliant'' powers of $u$, whereas any power of $u$ involving $\Delta$ is ``non-compliant.''  Then we notice that non-compliant powers cancel in the following partial sums:
 \eqnQq{PartialSums}{
  V_2 + V_3 &= -u^{\Delta_3+\Delta_4} \zeta(\Delta+\Delta_{13}) \zeta(-\Delta+\Delta_{24})  \cr
  V_{\tilde{2}} + V_{\tilde{3}} &= -u^{\Delta_1+\Delta_2} \zeta(\Delta-\Delta_{13})
      \zeta(-\Delta-\Delta_{24})  \cr
  V_{\underline{2}} + V_4 &= u^{\Delta_1+\Delta_2} \zeta(\Delta-\Delta_{13}) 
      \zeta(-\Delta_{13}-\Delta_{24}) \cr 
      &\quad + u^{\Delta_3+\Delta_4} \zeta(\Delta-\Delta_{13}) \zeta(\Delta_{13}+\Delta_{24})  \cr
  V_{\underline{\tilde{2}}} + V_{\tilde{4}} &= -u^{\Delta_1+\Delta_2}
      \zeta(-\Delta-\Delta_{13}) \zeta(-\Delta_{13}-\Delta_{24}) \cr
      &\quad - u^{\Delta_3+\Delta_4} \zeta(-\Delta-\Delta_{13}) \zeta(\Delta_{13}+\Delta_{24})\,.
 }
Finally, we can assemble the full exchange amplitude in the $(13)(24)$-channel 
 \eqnQq{GtAssembled}{
  {{\cal W}_\Delta^T \over W_0} &= V_1 + V_2 + V_3 + V_{\tilde{1}} + V_{\tilde{2}}  + V_{\tilde{3}} +  V_{\underline{2}} + V_4 \cr 
  &\quad + V_{\underline{\tilde{2}}} + V_{\tilde{4}}  \cr
   &= \beta^{(\Delta+\Delta_{13},\Delta-\Delta_{13})} \beta^{(-\Delta_{13}-\Delta_{24},\Delta+\Delta_{24})} u^{\Delta_1+\Delta_2} \cr
   &\quad + \beta^{(\Delta+\Delta_{13},\Delta-\Delta_{13})}  \beta^{(\Delta_{13}+\Delta_{24},\Delta-\Delta_{24})} u^{\Delta_3+\Delta_4}.
 }
This proves \eno{GeodesicConfWaveT}. Despite appearances, ${\cal W}_\Delta^T$ is symmetric under $1 \leftrightarrow 3$, $2 \leftrightarrow 4$ as well as $1 \leftrightarrow 2$, $3 \leftrightarrow 4$.  The amplitude with the exchange in the $(14)(23)$-channel is obtained by swapping $3 \leftrightarrow 4$ in ${\cal W}_\Delta^T$:
 \eqnQq{GuSwapped}{
  {{\cal W}_\Delta^U \over W_0} &= \beta^{(\Delta+\Delta_{14},\Delta-\Delta_{14})} \beta^{(-\Delta_{14}-\Delta_{23},\Delta+\Delta_{23})} u^{\Delta_1+\Delta_2} \cr
   &\quad + \beta^{(\Delta+\Delta_{14},\Delta-\Delta_{14})}  \beta^{(\Delta_{14}+\Delta_{23},\Delta-\Delta_{23})} u^{\Delta_3+\Delta_4}. 
 }

\bibliographystyle{ssg}
\bibliography{geodesic} 
\end{document}